\documentclass[a4paper, 10pt]{article}

\usepackage{jheppub}

\usepackage{amsthm}

\usepackage[numbers,sort&compress]{natbib}
\usepackage{dsfont}
\usepackage{slashed}
\usepackage{color}
\usepackage{empheq}

\usepackage{adjustbox}

\usepackage{bibgerm}

\usepackage{soul}

\usepackage[english]{babel}

\usepackage[utf8x]{inputenc}

\usepackage{graphicx,epsfig}
\usepackage{pstricks}

\usepackage{tikz}
\usepackage{tikz-feynman}

\usepackage{bashful}

\usepackage{subfigure}

\usepackage{setspace}

\usepackage{booktabs}

\usepackage{float}

\usepackage{nextpage}

\usepackage{pdfpages}

\usepackage{hypcap}

\usepackage[small,bf]{caption}

\usepackage{longtable}

\usepackage{fancyhdr}

\usepackage[makeroom]{cancel}

\usepackage{microtype}

\usepackage{xparse}
\usepackage{etoolbox}

\usepackage[subfigure]{tocloft}

\usepackage{listings}

\newcommand{\p}{\partial}

\newcommand{\<}{\langle}
\renewcommand{\>}{\rangle}

\renewcommand{\O}{\mathcal{O}}

\newcommand{\A}{\mathcal{A}}

\newcommand{\D}{\mathcal{D}}

\renewcommand{\L}{\mathcal{L}}

\newcommand{\E}{\mathcal{E}}
\newcommand{\nn}{\nonumber\\}

\newcommand{\q}{\mathsf{q}}

\newcommand{\msbar}{$\overline{\text{MS}}$}

\newcommand{\hc}{\mathrm{h.c.}}

\NewDocumentCommand{\lwc}{ m m O{} o }{
	L^{\ifblank{#3}{}{#3,}#2 }_{\IfNoValueTF{#4}{#1}{\substack{#1\\#4}}}
}

\makeatletter
  \def\my@tag@font{\normalsize}
  \def\maketag@@@#1{\hbox{\m@th\normalfont\my@tag@font#1}}
  \let\amsmath@eqref\eqref
  \renewcommand\eqref[1]{{\let\my@tag@font\relax\amsmath@eqref{#1}}}
\makeatother

\newcommand{\roverline}[1]{\mathpalette\doroverline{#1}}
\newcommand{\doroverline}[2]{\overline{#1#2}}

\makeatletter
\renewcommand\paragraph{\@startsection{paragraph}{4}{\z@}%
  {-3.25ex\@plus -1ex \@minus -.2ex}%
  {1.5ex \@plus .2ex}%
  {\normalfont\normalsize\bfseries}}
\makeatother

\cftsetindents{section}{0em}{2em}
\cftsetindents{subsection}{2em}{3em}
\cftsetindents{subsubsection}{5em}{4em}

\allowdisplaybreaks[1]

\preprint{
\mbox{}\hfill{} PSI-PR-24-30 \\
\mbox{}\hfill{} ZU-TH 66/24
}

\title{\boldmath Renormalization-group equations of the LEFT at two loops: dimension-five effects}

\author{Luca Naterop,}

\author{Peter Stoffer}

\emailAdd{luca.naterop@physik.uzh.ch}
\emailAdd{stoffer@physik.uzh.ch}

\affiliation{Physik-Institut, Universit\"at Z\"urich, Winterthurerstrasse 190, 8057 Z\"urich, Switzerland}
\affiliation{PSI Center for Neutron and Muon Sciences, 5232 Villigen PSI, Switzerland}

\abstract{
	We present the first part of a systematic calculation of the two-loop anomalous dimensions in the low-energy effective field theory (LEFT): the effects at dimension five in the power counting. Our calculation is performed in a basis with generic mass matrices. We employ the algebraically consistent 't Hooft--Veltman scheme for $\gamma_5$ and we correct for evanescent as well as chiral-symmetry-breaking effects by including the appropriate finite counterterms. We also provide results for the $CP$-even sector in a scheme that coincides with naive dimensional regularization. We discuss two methods to avoid the explicit construction of gauge-variant operators, which in principle are needed for the cancellation of sub-divergences, even in the background-field method. The two methods are consistent with each other and with existing partial results. Our work is a further step towards a complete EFT framework for physics beyond the Standard Model at next-to-leading-logarithmic accuracy.
}

\numberwithin{equation}{section}

\begin{document}

	\maketitle


\tikzset{every picture/.style={line width=0.7pt}}

\tikzset{
    cross/.pic = {
    \draw[rotate = 45] (-#1,0) -- (#1,0);
    \draw[rotate = 45] (0,-#1) -- (0, #1);
    }
}

\newcommand{\sunsetdiag}{
    \begin{tikzpicture}[scale=0.3,anchor=base,baseline=-0.6ex]
        \draw (0,0) circle (1.5);
        \draw (-1.5,0) -- (1.5,0);  
        \node at (0,+1.7) {\scriptsize 1};  
        \node at (0,+0.2) {\scriptsize 2};  
        \node at (0,-1.3) {\scriptsize 3};  
    \end{tikzpicture}
}

\newcommand{\sunsetdiagnonums}{
    \begin{tikzpicture}[scale=0.3,anchor=base,baseline=-0.6ex]
        \draw (0,0) circle (1.5);
        \draw (-1.5,0) -- (1.5,0);  
    \end{tikzpicture}
}

\newcommand{\infdiag}{
    \begin{tikzpicture}[scale=0.3,anchor=base,baseline=-0.6ex]
        \draw (-1,0) circle (1);
        \draw (1,0) circle (1);
        \node at (-1,+1.3) {\scriptsize 1};
        \node at (+1,+1.3) {\scriptsize 2};
    \end{tikzpicture}
}

\newcommand{\ctdiagOne}{
    \begin{tikzpicture}[scale=0.3,anchor=base,baseline=-0.6ex]
        \draw (0,0) circle (1);
        \path (1,0) pic {cross=3pt};
        \node at (0,+1.3) {\scriptsize 1};
    \end{tikzpicture}    
}

\newcommand{\ctdiag}{
    \begin{tikzpicture}[scale=0.3,anchor=base,baseline=-0.6ex]
        \draw (0,0) circle (1);
        \path (1,0) pic {cross=3pt};
    \end{tikzpicture}    
}

\newcommand{\ctdiagTwo}{
    \begin{tikzpicture}[scale=0.3,anchor=base,baseline=-0.6ex]
        \draw (0,0) circle (1);
        \path (-1,0) pic {cross=3pt};
        \node at (0,+1.3) {\scriptsize 2};
    \end{tikzpicture}    
}

\newcommand{\ctdiagOneAlt}{
    \begin{tikzpicture}[scale=0.3,anchor=base,baseline=-0.6ex]
        \draw (0,0) circle (1);
        \path (0,-1) pic {cross=3pt};
        \node at (0,+1.3) {\scriptsize 1};
    \end{tikzpicture}    
}

\newcommand{\ctdiagThreeAlt}{
    \begin{tikzpicture}[scale=0.3,anchor=base,baseline=-0.6ex]
        \draw (0,0) circle (1);
        \path (0,+1) pic {cross=3pt};
        \node at (0,-1.8) {\scriptsize 3};
    \end{tikzpicture}    
}
\newcommand{\overallct}{
    \begin{tikzpicture}[scale=0.3,anchor=base,baseline=-0.6ex]
        \path (0,0) pic {cross=3pt};
    \end{tikzpicture}    
}

\tikzset{every picture/.style={line width=0.9pt}}

\newcommand{\diagtitle}[1]{
  \node[above,font=\scriptsize \bfseries] at (current bounding box.north) {\texttt{#1}};
}

\newcommand{\base}{
    \begin{tikzpicture}
        \begin{feynman}
          \vertex (a);
          \vertex [right=1cm of a] (b);
          \diagram* {
            (a) -- [gluon] (b),
          };
        \end{feynman}
    \end{tikzpicture}
}

\newcommand{\gluondiagDA}{
    \begin{tikzpicture}
        \begin{feynman}
          \vertex (a);
          \vertex [right=0.7cm of a] (b);
          \vertex [right=1.3cm of b] (c);
          \vertex [right=0.7cm of c] (d);
          \diagram* {
            (a) -- [gluon] (b),
            (b) -- [gluon] (c),
            (b) -- [gluon, half right] (c),
            (c) -- [gluon, half right] (b),
            (c) -- [gluon] (d),
          };
        \end{feynman}
        \diagtitle{1}
    \end{tikzpicture}
}

\newcommand{\gluondiagDB}{
    \begin{tikzpicture}
        \begin{feynman}
          \vertex (a);
          \vertex [right=0.7cm of a] (b);
          \vertex [right=1.3cm of b] (c);
          \vertex [right=0.7cm of c] (d);
          \diagram* {
            (a) -- [gluon] (b),
            (b) -- [gluon] (c),
            (b) -- [fermion, half left] (c),
            (c) -- [fermion, half left] (b),
            (c) -- [gluon] (d),
          };
        \end{feynman}
        \diagtitle{2}
    \end{tikzpicture}
}

\newcommand{\gluondiagDC}{
    \begin{tikzpicture}
        \begin{feynman}
          \vertex (a);
          \vertex [right=0.7cm of a] (b);
          \vertex [right=1.3cm of b] (c);
          \vertex [right=0.7cm of c] (d);
          \diagram* {
            (a) -- [gluon] (b),
            (b) -- [gluon] (c),
            (b) -- [charged scalar, half left] (c),
            (c) -- [charged scalar, half left] (b),
            (c) -- [gluon] (d),
          };
        \end{feynman}
        \diagtitle{3}
    \end{tikzpicture}
}

\newcommand{\gluondiagEA}{
    \begin{tikzpicture}
        \begin{feynman}
          \vertex (a);
          \vertex [right=1.2cm of a] (b);
          \vertex [right=1.2cm of b] (c);
          \vertex [right=0.6cm of a] (a1);
          \vertex [left=0.6cm of c] (c1);
          \vertex [above=1cm of a1] (o1);
          \vertex [above=1cm of c1] (o2);
          \diagram* {
            (a) -- [gluon] (b),
            (b) -- [gluon] (c),
            (o1) -- [gluon, quarter right] (b),
            (b) -- [gluon, quarter right] (o2),
            (o1) -- [gluon, quarter right] (o2),
            (o2) -- [gluon, quarter right] (o1),
          };
        \end{feynman}
        \diagtitle{4}
    \end{tikzpicture}
}

\newcommand{\gluondiagEB}{
    \begin{tikzpicture}
        \begin{feynman}
          \vertex (a);
          \vertex [right=1.2cm of a] (b);
          \vertex [right=1.2cm of b] (c);
          \vertex [right=0.6cm of a] (a1);
          \vertex [left=0.6cm of c] (c1);
          \vertex [above=1cm of a1] (o1);
          \vertex [above=1cm of c1] (o2);
          \diagram* {
            (a) -- [gluon] (b),
            (b) -- [gluon] (c),
            (b) -- [fermion, quarter left] (o1),
            (o2) -- [fermion, quarter left] (b),
            (o1) -- [fermion, quarter left] (o2),
            (o1) -- [gluon, quarter right] (o2),
          };
        \end{feynman}
        \diagtitle{5}
    \end{tikzpicture}
}

\newcommand{\gluondiagEC}{
    \begin{tikzpicture}
        \begin{feynman}
          \vertex (a);
          \vertex [right=1.2cm of a] (b);
          \vertex [right=1.2cm of b] (c);
          \vertex [right=0.6cm of a] (a1);
          \vertex [left=0.6cm of c] (c1);
          \vertex [above=1cm of a1] (o1);
          \vertex [above=1cm of c1] (o2);
          \diagram* {
            (a) -- [gluon] (b),
            (b) -- [gluon] (c),
            (b) -- [charged scalar, quarter left] (o1),
            (o2) -- [charged scalar, quarter left] (b),
            (o1) -- [charged scalar, quarter left] (o2),
            (o1) -- [gluon, quarter right] (o2),
          };
        \end{feynman}
        \diagtitle{6}
    \end{tikzpicture}
}

\newcommand{\gluondiagED}{
    \begin{tikzpicture}
        \begin{feynman}
          \vertex (a);
          \vertex [right=1.2cm of a] (b);
          \vertex [right=1.2cm of b] (c);
          \vertex [right=0.6cm of a] (a1);
          \vertex [left=0.6cm of c] (c1);
          \vertex [above=1cm of a1] (o1);
          \vertex [above=1cm of c1] (o2);
          \diagram* {
            (a) -- [gluon] (b),
            (b) -- [gluon] (c),
            (o1) -- [gluon, quarter right] (b),
            (b) -- [gluon, quarter right] (o2),
            (o1) -- [fermion, quarter left] (o2),
            (o2) -- [fermion, quarter left] (o1),
          };
        \end{feynman}
        \diagtitle{7}
    \end{tikzpicture}
}

\newcommand{\gluondiagEE}{
    \begin{tikzpicture}
        \begin{feynman}
          \vertex (a);
          \vertex [right=1.2cm of a] (b);
          \vertex [right=1.2cm of b] (c);
          \vertex [right=0.6cm of a] (a1);
          \vertex [left=0.6cm of c] (c1);
          \vertex [above=1cm of a1] (o1);
          \vertex [above=1cm of c1] (o2);
          \diagram* {
            (a) -- [gluon] (b),
            (b) -- [gluon] (c),
            (o1) -- [gluon, quarter right] (b),
            (b) -- [gluon, quarter right] (o2),
            (o1) -- [charged scalar, quarter left] (o2),
            (o2) -- [charged scalar, quarter left] (o1),
          };
        \end{feynman}
        \diagtitle{8}
    \end{tikzpicture}
}

\newcommand{\gluondiagFA}{
    \begin{tikzpicture}
        \begin{feynman}
          \vertex (a);
          \vertex [right=0.6cm of a] (b);
          \vertex [right=1.2cm of b] (c);
          \vertex [above=0.6cm of b] (o1);
          \vertex [right=0.6cm of o1] (o);
          \vertex [below=0.6cm of b] (u1);
          \vertex [right=0.6cm of u1] (u);
          \vertex [right=0.6cm of c] (d);
          \diagram* {
            (a) -- [gluon] (b),
            (o) -- [gluon, quarter right] (b),
            (c) -- [gluon, quarter right] (o),
            (b) -- [gluon, quarter right] (u),
            (u) -- [gluon, quarter right] (c),
            (c) -- [gluon] (d),
            (o) -- [gluon] (u),
          };
        \end{feynman}
        \diagtitle{9}
    \end{tikzpicture}
}

\newcommand{\gluondiagFB}{
    \begin{tikzpicture}
        \begin{feynman}
          \vertex (a);
          \vertex [right=0.6cm of a] (b);
          \vertex [right=1.2cm of b] (c);
          \vertex [above=0.6cm of b] (o1);
          \vertex [right=0.6cm of o1] (o);
          \vertex [below=0.6cm of b] (u1);
          \vertex [right=0.6cm of u1] (u);
          \vertex [right=0.6cm of c] (d);
          \diagram* {
            (a) -- [gluon] (b),
            (b) -- [fermion, quarter left] (o),
            (o) -- [fermion, quarter left] (c),
            (u) -- [fermion, quarter left] (b),
            (c) -- [fermion, quarter left] (u),
            (c) -- [gluon] (d),
            (o) -- [gluon] (u),
          };
        \end{feynman}
        \diagtitle{10}
    \end{tikzpicture}
}

\newcommand{\gluondiagFC}{
    \begin{tikzpicture}
        \begin{feynman}
          \vertex (a);
          \vertex [right=0.6cm of a] (b);
          \vertex [right=1.2cm of b] (c);
          \vertex [above=0.6cm of b] (o1);
          \vertex [right=0.6cm of o1] (o);
          \vertex [below=0.6cm of b] (u1);
          \vertex [right=0.6cm of u1] (u);
          \vertex [right=0.6cm of c] (d);
          \diagram* {
            (a) -- [gluon] (b),
            (b) -- [charged scalar, quarter left] (o),
            (o) -- [charged scalar, quarter left] (c),
            (u) -- [charged scalar, quarter left] (b),
            (c) -- [charged scalar, quarter left] (u),
            (c) -- [gluon] (d),
            (o) -- [gluon] (u),
          };
        \end{feynman}
        \diagtitle{11}
    \end{tikzpicture}
}

\newcommand{\gluondiagFD}{
    \begin{tikzpicture}
        \begin{feynman}
          \vertex (a);
          \vertex [right=0.6cm of a] (b);
          \vertex [right=1.2cm of b] (c);
          \vertex [above=0.6cm of b] (o1);
          \vertex [right=0.6cm of o1] (o);
          \vertex [below=0.6cm of b] (u1);
          \vertex [right=0.6cm of u1] (u);
          \vertex [right=0.6cm of c] (d);
          \diagram* {
            (a) -- [gluon] (b),
            (b) -- [fermion, quarter left] (o),
            (c) -- [gluon, quarter right] (o),
            (u) -- [fermion, quarter left] (b),
            (u) -- [gluon, quarter right] (c),
            (c) -- [gluon] (d),
            (o) -- [fermion] (u),
          };
        \end{feynman}
        \diagtitle{12}
    \end{tikzpicture}
}

\newcommand{\gluondiagFE}{
    \begin{tikzpicture}
        \begin{feynman}
          \vertex (a);
          \vertex [right=0.6cm of a] (b);
          \vertex [right=1.2cm of b] (c);
          \vertex [above=0.6cm of b] (o1);
          \vertex [right=0.6cm of o1] (o);
          \vertex [below=0.6cm of b] (u1);
          \vertex [right=0.6cm of u1] (u);
          \vertex [right=0.6cm of c] (d);
          \diagram* {
            (a) -- [gluon] (b),
            (b) -- [charged scalar, quarter left] (o),
            (c) -- [gluon, quarter right] (o),
            (u) -- [charged scalar, quarter left] (b),
            (u) -- [gluon, quarter right] (c),
            (c) -- [gluon] (d),
            (o) -- [charged scalar] (u),
          };
        \end{feynman}
        \diagtitle{13}
    \end{tikzpicture}
}

\newcommand{\gluondiagGA}{
    \begin{tikzpicture}
        \begin{feynman}
          \vertex (a);
          \vertex [right=0.6cm of a] (b);
          \vertex [right=1.2cm of b] (c);
          \vertex [right=0.6cm of c] (d);
          \vertex [above=0.7cm of b] (u1);
          \vertex [right=0.2cm of u1] (u2);
          \vertex [right=0.8cm of u2] (u3);
          \diagram* {
            (a) -- [gluon] (b),
            (b) -- [gluon] (c),
            (c) -- [gluon] (d),
            (u2) -- [gluon, quarter right] (b),
            (u3) -- [gluon, half right] (u2),
            (u2) -- [gluon, half right] (u3),
            (c) -- [gluon, quarter right] (u3),
          };
        \end{feynman}
        \diagtitle{14}
    \end{tikzpicture}
}

\newcommand{\gluondiagGB}{
    \begin{tikzpicture}
        \begin{feynman}
          \vertex (a);
          \vertex [right=0.6cm of a] (b);
          \vertex [right=1.2cm of b] (c);
          \vertex [right=0.6cm of c] (d);
          \vertex [above=0.7cm of b] (u1);
          \vertex [right=0.2cm of u1] (u2);
          \vertex [right=0.8cm of u2] (u3);
          \diagram* {
            (a) -- [gluon] (b),
            (b) -- [gluon] (c),
            (c) -- [gluon] (d),
            (u2) -- [gluon, quarter right] (b),
            (u2) -- [fermion, half left] (u3),
            (u3) -- [fermion, half left] (u2),
            (c) -- [gluon, quarter right] (u3),
          };
        \end{feynman}
        \diagtitle{15}
    \end{tikzpicture}
}

\newcommand{\gluondiagGC}{
    \begin{tikzpicture}
        \begin{feynman}
          \vertex (a);
          \vertex [right=0.6cm of a] (b);
          \vertex [right=1.2cm of b] (c);
          \vertex [right=0.6cm of c] (d);
          \vertex [above=0.7cm of b] (u1);
          \vertex [right=0.2cm of u1] (u2);
          \vertex [right=0.8cm of u2] (u3);
          \diagram* {
            (a) -- [gluon] (b),
            (b) -- [gluon] (c),
            (c) -- [gluon] (d),
            (u2) -- [gluon, quarter right] (b),
            (u2) -- [charged scalar, half left] (u3),
            (u3) -- [charged scalar, half left] (u2),
            (c) -- [gluon, quarter right] (u3),
          };
        \end{feynman}
        \diagtitle{16}
    \end{tikzpicture}
}

\newcommand{\gluondiagGD}{
    \begin{tikzpicture}
        \begin{feynman}
          \vertex (a);
          \vertex [right=0.6cm of a] (b);
          \vertex [right=1.2cm of b] (c);
          \vertex [right=0.6cm of c] (d);
          \vertex [above=0.7cm of b] (u1);
          \vertex [right=0.2cm of u1] (u2);
          \vertex [right=0.8cm of u2] (u3);
          \diagram* {
            (a) -- [gluon] (b),
            (c) -- [fermion] (b),
            (c) -- [gluon] (d),
            (b) -- [fermion, quarter left] (u2),
            (u2) -- [fermion, half left] (u3),
            (u2) -- [gluon, half right] (u3),
            (u3) -- [fermion, quarter left] (c),
          };
        \end{feynman}
        \diagtitle{17}
    \end{tikzpicture}
}

\newcommand{\gluondiagGE}{
    \begin{tikzpicture}
        \begin{feynman}
          \vertex (a);
          \vertex [right=0.6cm of a] (b);
          \vertex [right=1.2cm of b] (c);
          \vertex [right=0.6cm of c] (d);
          \vertex [above=0.7cm of b] (u1);
          \vertex [right=0.2cm of u1] (u2);
          \vertex [right=0.8cm of u2] (u3);
          \diagram* {
            (a) -- [gluon] (b),
            (c) -- [charged scalar] (b),
            (c) -- [gluon] (d),
            (b) -- [charged scalar, quarter left] (u2),
            (u2) -- [charged scalar, half left] (u3),
            (u2) -- [gluon, half right] (u3),
            (u3) -- [charged scalar, quarter left] (c),
          };
        \end{feynman}
        \diagtitle{18}
    \end{tikzpicture}
}

\newcommand{\gluondiagHA}{
    \begin{tikzpicture}
        \begin{feynman}
          \vertex (a);
          \vertex [right=0.6cm of a] (b);
          \vertex [right=1.2cm of b] (c);
          \vertex [above=0.8cm of b] (u1);
          \vertex [right=0.6cm of u1] (u2);
          \diagram* {
            (a) -- [gluon] (b),
            (u2) -- [gluon, quarter right] (b),
            (u2) -- [gluon, quarter right] (c),
            (c) -- [gluon, quarter right] (u2),
            (b) -- [gluon] (c),
            (c) -- [gluon] (d),
          };
        \end{feynman}
        \diagtitle{19}
    \end{tikzpicture}
}

\newcommand{\gluondiagHB}{
    \begin{tikzpicture}
        \begin{feynman}
          \vertex (a);
          \vertex [right=0.6cm of a] (b);
          \vertex [right=1.2cm of b] (c);
          \vertex [above=0.8cm of b] (u1);
          \vertex [right=0.6cm of u1] (u2);
          \diagram* {
            (a) -- [gluon] (b),
            (u2) -- [gluon, quarter right] (b),
            (u2) -- [fermion, quarter left] (c),
            (c) -- [fermion, quarter left] (u2),
            (b) -- [gluon] (c),
            (c) -- [gluon] (d),
          };
        \end{feynman}
        \diagtitle{20}
    \end{tikzpicture}
}

\newcommand{\gluondiagHC}{
    \begin{tikzpicture}
        \begin{feynman}
          \vertex (a);
          \vertex [right=0.6cm of a] (b);
          \vertex [right=1.2cm of b] (c);
          \vertex [above=0.8cm of b] (u1);
          \vertex [right=0.6cm of u1] (u2);
          \diagram* {
            (a) -- [gluon] (b),
            (u2) -- [gluon, quarter right] (b),
            (u2) -- [charged scalar, quarter left] (c),
            (c) -- [charged scalar, quarter left] (u2),
            (b) -- [gluon] (c),
            (c) -- [gluon] (d),
          };
        \end{feynman}
        \diagtitle{21}
    \end{tikzpicture}
}

\newcommand{\gluondiagHD}{
    \begin{tikzpicture}
        \begin{feynman}
          \vertex (a);
          \vertex [right=0.6cm of a] (b);
          \vertex [right=1.2cm of b] (c);
          \vertex [above=0.8cm of b] (u1);
          \vertex [right=0.6cm of u1] (u2);
          \diagram* {
            (a) -- [gluon] (b),
            (b) -- [fermion, quarter left] (u2),
            (u2) -- [gluon, quarter right] (c),
            (u2) -- [fermion, quarter left] (c),
            (c) -- [fermion] (b),
            (c) -- [gluon] (d),
          };
        \end{feynman}
        \diagtitle{22}
    \end{tikzpicture}
}
\newcommand{\gluondiagHE}{
    \begin{tikzpicture}
        \begin{feynman}
          \vertex (a);
          \vertex [right=0.6cm of a] (b);
          \vertex [right=1.2cm of b] (c);
          \vertex [above=0.8cm of b] (u1);
          \vertex [right=0.6cm of u1] (u2);
          \diagram* {
            (a) -- [gluon] (b),
            (b) -- [charged scalar, quarter left] (u2),
            (u2) -- [gluon, quarter right] (c),
            (u2) -- [charged scalar, quarter left] (c),
            (c) -- [charged scalar] (b),
            (c) -- [gluon] (d),
          };
        \end{feynman}
        \diagtitle{23}
    \end{tikzpicture}
}

\newcommand{\gluondiagAA}{
    \begin{tikzpicture}
        \begin{feynman}
          \vertex (a);
          \vertex [right=0.8cm of a] (b);
          \vertex [right=0.8cm of b] (c);
          \vertex [above=0.8cm of b] (o);
          \diagram* {
            (a) -- [gluon] (b),
            (b) -- [gluon] (c),
            (b) -- [gluon, half right] (o),
            (o) -- [gluon, half right] (b),
          };
          \draw[gluon] (o) arc [start angle=-90, end angle=+270, radius=0.4cm];
        \end{feynman}
        \diagtitle{24}
    \end{tikzpicture}
}

\newcommand{\gluondiagAB}{
    \begin{tikzpicture}
        \begin{feynman}
          \vertex (a);
          \vertex [right=0.8cm of a] (b);
          \vertex [right=0.8cm of b] (c);
          \vertex [above=0.8cm of b] (o);
          \diagram* {
            (a) -- [gluon] (b),
            (b) -- [gluon] (c),
            (b) -- [gluon, half right] (o),
            (o) -- [gluon, half right] (b),
          };
          \draw[fermion] (o) arc [start angle=270, end angle=-90, radius=0.4cm];
        \end{feynman}
        \diagtitle{25}
    \end{tikzpicture}
}

\newcommand{\gluondiagAC}{
    \begin{tikzpicture}
        \begin{feynman}
          \vertex (a);
          \vertex [right=0.8cm of a] (b);
          \vertex [right=0.8cm of b] (c);
          \vertex [above=0.8cm of b] (o);
          \diagram* {
            (a) -- [gluon] (b),
            (b) -- [gluon] (c),
            (b) -- [fermion, half left] (o),
            (o) -- [fermion, half left] (b),
          };
          \draw[gluon] (o) arc [start angle=-90, end angle=270, radius=0.4cm];
        \end{feynman}
        \diagtitle{26}
    \end{tikzpicture}
}

\newcommand{\gluondiagAD}{
    \begin{tikzpicture}
        \begin{feynman}
          \vertex (a);
          \vertex [right=0.8cm of a] (b);
          \vertex [right=0.8cm of b] (c);
          \vertex [above=0.8cm of b] (o);
          \diagram* {
            (a) -- [gluon] (b),
            (b) -- [gluon] (c),
            (b) -- [charged scalar, half left] (o),
            (o) -- [charged scalar, half left] (b),
          };
          \draw[gluon] (o) arc [start angle=-90, end angle=270, radius=0.4cm];
        \end{feynman}
        \diagtitle{27}
    \end{tikzpicture}
}

\newcommand{\gluondiagAE}{
    \begin{tikzpicture}
        \begin{feynman}
          \vertex (a);
          \vertex [right=0.8cm of a] (b);
          \vertex [right=0.8cm of b] (c);
          \vertex [above=0.8cm of b] (o);
          \diagram* {
            (a) -- [gluon] (b),
            (b) -- [gluon] (c),
          };
          \draw[gluon] (b) arc [start angle=-90, end angle=270, radius=0.4cm];
          \draw[charged scalar] (o) arc [start angle=270, end angle=-90, radius=0.4cm];
        \end{feynman}
        \diagtitle{28}
    \end{tikzpicture}
}

\newcommand{\gluondiagBA}{
    \begin{tikzpicture}
        \begin{feynman}
          \vertex (a);
          \vertex [right=0.7cm of a] (b);
          \vertex [right=1cm of b] (c);
          \vertex [right=0.7cm of c] (d);
          \vertex [above right=0.65cm of b] (u1);
          \vertex [above=1.0cm of u1] (u2);
          \diagram* {
            (a) -- [gluon] (b),
            (b) -- [gluon, half right] (c),
            (c) -- [gluon, quarter right] (u1),
            (u1) -- [gluon, quarter right] (b),
            (c) -- [gluon] (d),
          };
          \draw[gluon] (u1) arc [start angle=-90, end angle=270, radius=0.4cm];
        \end{feynman}
        \diagtitle{29}
    \end{tikzpicture}
}

\newcommand{\gluondiagBB}{
    \begin{tikzpicture}
        \begin{feynman}
          \vertex (a);
          \vertex [right=0.7cm of a] (b);
          \vertex [right=1cm of b] (c);
          \vertex [right=0.7cm of c] (d);
          \vertex [above right=0.65cm of b] (u1);
          \vertex [above=1.0cm of u1] (u2);
          \diagram* {
            (a) -- [gluon] (b),
            (b) -- [fermion, quarter left] (u1),
            (u1) -- [fermion, quarter left] (c),
            (c) -- [fermion, half left] (b),
            (c) -- [gluon] (d),
          };
          \draw[gluon] (u1) arc [start angle=-90, end angle=270, radius=0.4cm];
        \end{feynman}
        \diagtitle{30}
    \end{tikzpicture}
}

\newcommand{\gluondiagBC}{
    \begin{tikzpicture}
        \begin{feynman}
          \vertex (a);
          \vertex [right=0.7cm of a] (b);
          \vertex [right=1cm of b] (c);
          \vertex [right=0.7cm of c] (d);
          \vertex [above right=0.65cm of b] (u1);
          \vertex [above=1.0cm of u1] (u2);
          \diagram* {
            (a) -- [gluon] (b),
            (b) -- [gluon, half right] (c),
            (c) -- [gluon, quarter right] (u1),
            (u1) -- [gluon, quarter right] (b),
            (c) -- [gluon] (d),
          };
          \draw[fermion] (u1) arc [start angle=270, end angle=-90, radius=0.4cm];
        \end{feynman}
        \diagtitle{31}
    \end{tikzpicture}
}

\newcommand{\gluondiagBD}{
    \begin{tikzpicture}
        \begin{feynman}
          \vertex (a);
          \vertex [right=0.7cm of a] (b);
          \vertex [right=1cm of b] (c);
          \vertex [right=0.7cm of c] (d);
          \vertex [above right=0.65cm of b] (u1);
          \vertex [above=1.0cm of u1] (u2);
          \diagram* {
            (a) -- [gluon] (b),
            (b) -- [gluon, half right] (c),
            (c) -- [gluon, quarter right] (u1),
            (u1) -- [gluon, quarter right] (b),
            (c) -- [gluon] (d),
          };
          \draw[charged scalar] (u1) arc [start angle=270, end angle=-90, radius=0.4cm];
        \end{feynman}
        \diagtitle{32}
    \end{tikzpicture}
}

\newcommand{\gluondiagCA}{
    \begin{tikzpicture}
        \begin{feynman}
          \vertex (a);
          \vertex [right=0.7cm of a] (b);
          \vertex [right=0.8cm of b] (c);
          \vertex [right=0.8cm of c] (d);
          \vertex [right=0.7cm of d] (e);
          \diagram* {
            (b) -- [gluon] (a),
            (e) -- [gluon] (d),
          };
          \draw[gluon] (b) arc [start angle=-180, end angle=180, radius=0.4cm];
          \draw[gluon] (c) arc [start angle=-180, end angle=180, radius=0.4cm];
        \end{feynman}
        \diagtitle{33}
    \end{tikzpicture}
}

\newcommand{\gluondiagCB}{
    \begin{tikzpicture}
        \begin{feynman}
          \vertex (a);
          \vertex [right=0.7cm of a] (b);
          \vertex [right=0.8cm of b] (c);
          \vertex [right=0.8cm of c] (d);
          \vertex [right=0.7cm of d] (e);
          \diagram* {
            (a) -- [gluon] (b),
            (d) -- [gluon] (e),
            (c) -- [fermion, half left] (d),
            (d) -- [fermion, half left] (c),
          };
          \draw[gluon] (b) arc [start angle=-180, end angle=+180, radius=0.4cm];
        \end{feynman}
        \diagtitle{34}
    \end{tikzpicture}
}

\newcommand{\gluondiagCC}{
    \begin{tikzpicture}
        \begin{feynman}
          \vertex (a);
          \vertex [right=0.7cm of a] (b);
          \vertex [right=0.8cm of b] (c);
          \vertex [right=0.8cm of c] (d);
          \vertex [right=0.7cm of d] (e);
          \diagram* {
            (a) -- [gluon] (b),
            (d) -- [gluon] (e),
            (c) -- [charged scalar, half left] (d),
            (d) -- [charged scalar, half left] (c),
          };
          \draw[gluon] (b) arc [start angle=-180, end angle=+180, radius=0.4cm];
        \end{feynman}
        \diagtitle{35}
    \end{tikzpicture}
}


\section{Introduction}

Effective field theories (EFTs) are a key tool in many areas of theoretical physics: the restriction to the relevant degrees of freedom simplifies calculations that would be difficult or even impossible to perform based on the underlying ultraviolet (UV) theory. The explicit separation of scales also enables the resummation of large logarithms, thus improving perturbation theory. The interest in EFTs describing the low-energy effects of heavy particles beyond the Standard Model (SM) has considerably increased due to the absence of signals of new physics in direct collider searches. Under the assumption of linear realization of the electroweak symmetry, the appropriate EFT at energies above the electroweak scale is the SMEFT~\cite{Buchmuller:1985jz,Grzadkowski:2010es}. The low-energy EFT below the electroweak scale (LEFT) is obtained by integrating out the heavy SM particles.\footnote{The LEFT does not make any assumptions about electroweak symmetry breaking, hence it is also the correct low-energy theory in case of a nonlinear realization of the electroweak symmetry~\cite{Feruglio:1992wf,Grinstein:2007iv,Alonso:2012px,Buchalla:2013rka,Buchalla:2013eza,Gavela:2016bzc,Pich:2016lew}. In a linear realization there exist additional constraints on LEFT operators \cite{Karmakar:2024gla,Bause:2020auq,Bause:2021cna}, encoded in the matching conditions~\cite{Jenkins:2017jig,Dekens:2019ept}. } The operator bases of the SMEFT~\cite{Buchmuller:1985jz,Grzadkowski:2010es,Lehman:2014jma,Liao:2016hru,Murphy:2020rsh,Li:2020gnx,Liao:2020jmn,Harlander:2023psl} and LEFT~\cite{Jenkins:2017jig,Liao:2020zyx,Murphy:2020cly,Li:2020tsi} are known to high dimensions and the complete one-loop renormalization of the two theories up to dimension six was performed in Refs.~\cite{Jenkins:2013zja,Jenkins:2013wua,Alonso:2013hga,Jenkins:2017dyc}. The renormalization beyond one loop is partially known, for some sectors of the LEFT even up to four loops~\cite{Buras:1989xd,Buras:1991jm,Buras:1992tc,Ciuchini:1993vr,Ciuchini:1993fk,Buchalla:1995vs,Chetyrkin:1997gb,Buras:2000if,Bobeth:2003at,Gorbahn:2004my,Huber:2005ig,Gorbahn:2005sa,Czakon:2006ss,Aebischer:2017gaw,Panico:2018hal,Morell:2024aml,Aebischer:2025hsx}. Some results for the SMEFT renormalization beyond dimension six have been obtained as well~\cite{Liao:2016hru,Liao:2019tep,Chala:2021juk,Chala:2021pll,AccettulliHuber:2021uoa,DasBakshi:2022mwk,Helset:2022pde,DasBakshi:2023htx,Zhang:2023kvw,Chala:2023xjy,Zhang:2023ndw,Boughezal:2024zqa,Bakshi:2024wzz}. The status of SMEFT and LEFT was discussed in recent reviews~\cite{Brivio:2017vri,Isidori:2023pyp}.

In order to elevate these EFTs to next-to-leading-logarithmic (NLL) accuracy, matching and matrix elements at one loop and renormalization-group equations (RGEs) at two loops are required. The one-loop matching between SMEFT and LEFT was computed in Ref.~\cite{Dekens:2019ept} and automated tools enable the one-loop matching of UV models to SMEFT either diagrammatically or using functional methods~\cite{Carmona:2021xtq,Fuentes-Martin:2022jrf,Fuentes-Martin:2023ljp,Aebischer:2023nnv,Thomsen:2024abg}, while a broad effort is ongoing to compute the complete two-loop RGEs~\cite{deVries:2019nsu,Bern:2020ikv,Aebischer:2022anv,Fuentes-Martin:2022vvu,Aebischer:2023djt,Jenkins:2023rtg,Jenkins:2023bls,Naterop:2023dek,Fuentes-Martin:2023ljp,Aebischer:2024xnf,Manohar:2024xbh,DiNoi:2024ajj,Born:2024mgz,DiNoi:2023ygk,Fuentes-Martin:2024agf}. At NLL, scheme dependences start to show up, which need to cancel between finite one-loop terms in matching contributions and matrix elements and the two-loop RGEs when NLL resummation is performed. In Ref.~\cite{Naterop:2023dek}, we advocated the use of the 't~Hooft--Veltman (HV) scheme \cite{tHooft:1972tcz} for the LEFT starting at NLL. It is the only scheme proven to be algebraically consistent to all loop orders~\cite{Breitenlohner:1975hg,Breitenlohner:1976te,Breitenlohner:1977hr} and sometimes it is also called Breitenlohner--Maison/'t~Hooft--Veltman (BMHV) scheme.\footnote{The simpler naive dimensional regularization (NDR) scheme in general leads to ill-defined $\gamma_5$-odd traces, see Ref.~\cite{Jegerlehner:2000dz} for a review. As an alternative to the HV scheme, one could give up the cyclicity of the trace~\cite{Kreimer:1989ke,Korner:1991sx}, but we are not aware of a proof of the consistency of such a prescription that applies to non-renormalizable theories.} It comes with the difficulty of an extended evanescent sector and that the restoration of symmetries broken by the regulator requires finite counterterms~\cite{Schubert:1988ke,Ferrari:1994ct,Cornella:2022hkc,Belusca-Maito:2020ala,Belusca-Maito:2021lnk,Belusca-Maito:2023wah,Stockinger:2023ndm,OlgosoRuiz:2024dzq,Ebert:2024xpy}. Our NLL scheme for the LEFT avoids a spurious breaking of chiral symmetry and separates the physical and evanescent sectors by including such finite renormalizations~\cite{Naterop:2023dek}.

Although the computation of higher-order RGEs is well established, the completion of this program for SMEFT and LEFT is computationally demanding due to the large number of effective operators. In addition, the dimensional split in the HV scheme typically leads to a very large number of terms in intermediate results, necessitating an efficient algorithm and a high degree of automation for these calculations. Some aspects of the computations can be simplified by making use of the background-field method~\cite{Abbott:1980hw,Abbott:1983zw}, which in the LEFT preserves manifest gauge invariance. In the background-field method, gauge-variant nuisance operators appear in the renormalization of quantum-field sub-amplitudes as dictated by BRST symmetry~\cite{Dixon:1974ss,Kluberg-Stern:1975ebk,Joglekar:1975nu,Deans:1978wn,Falcioni:2022fdm}. For this reason, in previously used approaches the advantages of the background-field method were partially lost at higher loop orders and many calculations were instead performed in standard $R_\xi$ gauges. We discuss two methods that avoid the explicit construction of gauge-variant nuisance operators, even when computing off shell: the first is based on the well-established local $R$-operation, which automatically subtracts all sub-divergences from a given two-loop diagram. As an alternative, we introduce a variant of the infrared (IR) rearrangement, which separates UV and IR divergences without the need to introduce auxiliary-mass counterterms~\cite{Misiak:1994zw,Chetyrkin:1997fm,Gambino:2003zm,Gorbahn:2005sa,Lang:2020nnl,Stockinger:2023ndm}. When combined with the background-field method, we show that the correct renormalization of the non-redundant physical operators can be obtained from an off-shell calculation without computing counterterm diagrams with insertions of redundant operators. This method simplifies alternative approaches based on a global renormalization that require the explicit construction of gauge-variant operators and their insertion into counterterm diagrams. We compare this method with the local $R$-operation and find full agreement.

The article is structured as follows. In Sect.~\ref{sec:NuisanceOperators}, we discuss the role of nuisance operators in gauge theories, both gauge-invariant redundant operators that vanish by the equations of motion (EOM) and gauge-variant (class-IIb) nuisance operators, using the examples of QED and QCD augmented by dimension-five operators. In Sect.~\ref{sec:IRrearrangement}, we discuss two variants of the IR rearrangement and we show that the expansion of loop integrands allows us to ignore redundant operators. In Sect.~\ref{sec:LEFT}, we provide details on our calculation of the two-loop RGEs for the LEFT at dimension five. Explicit results for the two-loop counterterms of one-flavor QED and QCD with $CP$-even dimension-five operators are given in App.~\ref{sec:Counterterms}, whereas the RGEs for the full LEFT at dimension 5 in the HV scheme can be found in App.~\ref{sec:LEFTRGE}.


\section{Nuisance operators in gauge theories}
\label{sec:NuisanceOperators}

In this section, we discuss some well-known properties of so-called nuisance operators. These operators do not contribute to observables and hence they are redundant. In Sect.~\ref{sec:NuisanceGeneralities}, we review the different types of operators in gauge theories. In Sect.~\ref{sec:EOMOperators}, we illustrate the case of gauge-invariant operators that vanish by the EOM with an abelian example, recalling the connection to field redefinitions and the reason why the redundant operators do not contribute to the $S$-matrix. In Sect.~\ref{sec:ClassIIbNuisance}, we extend the discussion to BRST-exact gauge-variant operators. We will later use these properties to show how the counterterms of physical operators can be obtained from an off-shell calculation without insertion of redundant operators.

\subsection{Nuisance operators, the background-field method, and sub-divergences}
\label{sec:NuisanceGeneralities}

The renormalization of gauge theories in general requires the following operators~\cite{Dixon:1974ss,Kluberg-Stern:1975ebk,Joglekar:1975nu,Deans:1978wn}.
\begin{itemize}
	\item class I: {\em physical} operators, i.e., gauge-invariant operators that do not vanish by the EOM,
	\item class IIa: gauge-invariant {\em nuisance} operators that vanish by the EOM,
	\item class IIb: gauge-variant, BRST-exact {\em nuisance} operators.
\end{itemize}
The nuisance operators of class IIb can be constructed as BRST variations of operators with ghost number $-1$~\cite{Deans:1978wn}. They consist of gauge-variant operators that contain ghost terms or vanish by the EOM. In the background-field method~\cite{Abbott:1980hw,Abbott:1983zw}, the one-particle-irreducible (1PI) effective action can be computed without fixing the gauge of the background fields, hence manifest gauge invariance is preserved with respect to background-gauge transformations. Green's functions of background fields do not require class-IIb operators as overall counterterms. However, sub-diagrams are given by Green's functions of quantum fields: the cancellation of sub-divergences therefore in general still requires the introduction of class-IIb operators~\cite{Misiak:1994zw,Falcioni:2022fdm}. The explicit construction of these nuisance operators can be avoided when using the local $R$-operation~\cite{Chetyrkin:1982nn,Chetyrkin:1984xa,Smirnov:1985yck,Herzog:2017bjx}, which however typically leads to a large number of sub-diagrams that need to be computed.

In addition to using the local $R$-operation, in Sect.~\ref{sec:IRrearrangement} we will present a variant of the IR rearrangement that allows us to disregard class-II nuisance operators in sub-diagrams for the calculation of the two-loop counterterms in the physical sector. Although this procedure is based on the observation that nuisance operators do not contribute to the $S$-matrix, the calculation can be done off shell and does not require a transformation to the mass basis.

\subsection{Gauge-invariant redundant operators}
\label{sec:EOMOperators}

We consider an EFT consisting of single-flavor QED augmented by $CP$-even dimension-five operators,
\begin{align}
	\label{eq:QEDLagrangian}
	\L[\psi,J] &= -\frac{1}{4} F_{\mu\nu} F^{\mu\nu} + \bar \psi \, (i \slashed D-m) \psi  + L \, \bar \psi \,\sigma_{\mu \nu} F^{\mu \nu} \psi + R \, \bar \psi \, (i \slashed{D} - m)^2 \psi \nn
		&\quad + \bar J \psi + \bar \psi J + J_\mu A^\mu + \L_\mathrm{gf} \, ,
\end{align}
where the argument $\psi$ denotes collectively the fields $\psi$, $\bar\psi$, and $A_\mu$, and $J$ stands for the corresponding external sources $\bar J$, $J$, and $J_\mu$. We provide the two-loop counterterms and anomalous dimensions for Eq.~\eqref{eq:QEDLagrangian} in App.~\ref{sec:QEDDipoleResults}. The covariant derivative is $D_\mu = \p_\mu + i e \q A_\mu$, the field-strength tensor is given by $F_{\mu\nu} = \p_\mu A_\nu - \p_\nu A_\mu$, and the gauge-fixing Lagrangian is
\begin{equation}
	\L_\mathrm{gf} = - \frac{1}{2\xi} (\p^\mu A_\mu)^2 \, ,
\end{equation}
which does not lead to interaction vertices. In QED, ghosts decouple and can be ignored. Therefore, in the present case no gauge-variant class-IIb counterterms are generated. The class-IIa EOM operator with coefficient $R$ in Eq.~\eqref{eq:QEDLagrangian} is redundant, as it can be removed by a field redefinition.

Green's functions are obtained from the generating functional
\begin{equation}
	Z[J] = e^{i W[J]} = \int \D\psi \, \D\bar\psi \, \D A \, \exp\left\{ i \int d^Dx \, \L[\psi,J] \right\}
\end{equation}
by taking functional derivatives with respect to the sources,
\begin{equation}
	\< \psi(x_1) \bar \psi(x_2) A_\mu(x_3) \cdots \; \> = \frac{1}{Z[0]} \frac{-i \delta}{\delta \bar J(x_1)} \frac{i \delta}{\delta J(x_2)}  \frac{-i \delta}{\delta J^\mu(x_3)} \cdots \,Z[J] \, \bigg|_{J=0} \, . 
\end{equation}
Field redefinitions, such as those that remove redundant operators, simply reparametrize the path integral $Z[J]$, and thus leave $Z[J]$ manifestly invariant.\footnote{In dimensional regularization, the Jacobian is equal to identity for local field redefinitions.} Therefore, the Green's functions of the original fields can be computed with a redefined Lagrangian, provided that the transformation of the source terms is taken into account~\cite{Manohar:2018aog,Criado:2018sdb}. Explicitly, under a field redefinition
\begin{equation}
	\psi = F[\hat\psi] \, ,
\end{equation}
the Green's functions of the original fields computed with the original Lagrangian
\begin{equation}
	\< \psi \cdots e^{i S[\psi]} \> = \< F[\hat\psi] \cdots e^{i S[F[\hat\psi]]}\>
\end{equation}
are the same as Green's functions of the original fields computed with the redefined Lagrangian, which in terms of the new fields correspond to Green's functions of operators $F[\hat\psi]$.

Considering the QED example, we make a field redefinition to remove the redundant operator,
\begin{equation}
	\psi = F[\hat\psi] = \hat\psi - \frac{R}{2} (i \slashed D-m) \hat \psi,
\end{equation}
keeping track of the source terms,
\begin{align}
	\L[\psi,J] &= \L[F[\hat\psi],J] = \L'[\hat\psi,J] \nn
		&= \L[\hat \psi, J] ~ - R \, \hat{\bar \psi} \,(i \slashed D-m)^2 \hat \psi ~ - \frac{R}{2} \hat{\bar \psi} \, (i \slashed D-m) J  ~ - \frac{R}{2} \bar J \, (i \slashed D-m) \hat \psi + \O(\text{dim-6}) \, .
    \label{eq:byeredundant}
\end{align}
The redundant operator drops in the difference of the first two terms, but additional source terms have arisen. We can perform another field redefinition
\begin{equation}
	\hat \psi = \tilde \psi + \frac{R}{2} J \, ,
\end{equation}
which brings them back into canonical form, 
\begin{equation}
	\L''[\tilde\psi, J] = \L[\tilde \psi, J] ~ - R \, \tilde{ \bar \psi} \,(i \slashed D-m)^2 \tilde \psi + R \, \bar J J  + \O(\text{dim-6}) \, ,
	\label{eq:Ldoubleprime}
\end{equation}
but introduces a quadratic source term $R\, \bar J J$. When one takes functional derivatives with respect to the sources to calculate the two-point function in momentum space, this term gives a constant contribution proportional to $R$. Upon amputation of external legs, such a contribution is proportional to EOM terms $\slashed p - m$ and thus vanishes for on-shell external states. In an off-shell renormalization, it corresponds to an overall contact contribution of the EOM operator. 

$\L''$ must give the same off-shell Green's function as $\L$, since during the field redefinitions we have kept track of the source terms. When renormalizing the theory at some loop order $l$, a given subdivergence-subtracted connected graph has some remaining local divergence. The same divergence is found with either $\L$ or $\L''$. But $\L''$ has no EOM operator, only the $R\, \bar J J$ term, which has no effect on loop diagrams. Therefore, when using $\L''$ there are no insertions of the redundant operator into loop diagrams; there is only the contact term at tree level, which determines $R$ at $l$ loops.  

Putting everything together, Eq.~\eqref{eq:Ldoubleprime} states that, using field redefinitions, we can remove the redundant operator $R \,  \bar \psi \,(i \slashed D-m)^2 \psi$ at the cost of introducting a term $R \, \bar J J$, which does not enter loops. The procedure must leave $Z[J]$ (which generates all diagrams) and $W[J]$ (which generates connected diagrams) invariant, since the field redefinition is merely a change of variables in the path integral. Renormalization calculations are usually performed in terms of the 1PI effective action $\Gamma[J]$, which is the Legendre transform of $W[J]$. It turns out that in general $\Gamma[J]$ does change, since non-linear field redefinitions can turn a 1PI diagram into a one-particle-reducible diagram and vice versa. Nevertheless, we argue that leaving out redundant operators in $\Gamma[J]$ is still possible, see Sect.~\ref{sec:IRdimreg}. 

Having stated the general argument, we would like to explicitly test field-redefinition invariance for $W[J]$ in a toy calculation for connected diagrams. The result that we find is that, as expected, connected Green's functions agree when using $\L$ versus $\L''$ up to the two-loop level covered by our check. Starting at tree level, in the following we briefly discuss this check. Using crosses for external currents, with $\L''$ we find
\begin{align}
	\begin{tikzpicture}[baseline=-0.08cm]
        \begin{feynman}
          \vertex (a);
          \diagram* {
            a [crossed dot],
          };
        \end{feynman}
    \end{tikzpicture} \begin{tikzpicture}[baseline=-0.08cm]
        \begin{feynman}
          \vertex (a);
          \diagram* {
            a [crossed dot],
          };
        \end{feynman}
    \end{tikzpicture} = - i R \quad \xrightarrow{\text{amp.}} \quad i R (\slashed{p} - m)^2 \, ,
\end{align}
where the arrow means amputation, i.e. multiplication with $(\slashed{p}-m)/i$ from left and right. The $(-1)$ in the vertex rule of the contact term is due to the Grassmann algebra,
\begin{equation}
	\frac{\delta}{\delta \bar J}\frac{\delta}{\delta J} R  \bar J J = - R \, .
\end{equation}
The same result is found using $\L$, where at tree level we obtain
\begin{equation}
	\begin{tikzpicture}[baseline=-0.08cm]
        \begin{feynman}
          \vertex (a);
		  \vertex [right=1cm of a] (b);
		  \vertex [right=1cm of b] (c);
          \diagram* {
            a [crossed dot] -- [fermion] b [square dot] -- [fermion] c [crossed dot],
          };
        \end{feynman}
    \end{tikzpicture} \; = \; \frac{i}{\slashed p - m} i R (\slashed p - m)^2 \frac{i}{\slashed p - m} \quad \xrightarrow{\text{amp.}} \quad i R (\slashed{p} - m)^2 \, ,
\end{equation}
with the box denoting insertions of effective operators, which in this case means an insertion of $R \,  \bar \psi \,(i \slashed D-m)^2 \psi$. At one loop, the diagrams obtained from $\L''$ are 
\begin{equation}
	\begin{tikzpicture}[baseline=-0.08cm]
        \begin{feynman}
          \vertex (a);
		  \vertex [right=1cm of a] (b);
		  \vertex [right=1cm of b] (c);
		  \vertex [right=1cm of c] (d);
          \diagram* {
            a [crossed dot] -- [fermion] b  -- [fermion] c -- [fermion] d [crossed dot],
			(b) -- [photon, half left] (c)
          };
        \end{feynman}
    \end{tikzpicture} \quad + \quad 
	\begin{tikzpicture}[baseline=-0.08cm]
        \begin{feynman}
          \vertex (a);
		  \vertex [right=1cm of a] (b);
		  \vertex [right=1cm of b] (c);
		  \vertex [right=1cm of c] (d);
          \diagram* {
            a [crossed dot] -- [fermion] b [square dot]  -- [fermion] c -- [fermion] d [crossed dot],
			(b) -- [photon, half left] (c)
          };
        \end{feynman}
    \end{tikzpicture} \quad + \quad 
	\begin{tikzpicture}[baseline=-0.08cm]
        \begin{feynman}
          \vertex (a);
		  \vertex [right=1cm of a] (b);
		  \vertex [right=1cm of b] (c);
		  \vertex [right=1cm of c] (d);
          \diagram* {
            a [crossed dot] -- [fermion] b  -- [fermion] c [square dot] -- [fermion] d [crossed dot],
            (b) -- [photon, half left] (c)
          };
        \end{feynman}
    \end{tikzpicture} \;. 
\end{equation}
There are no two-point insertions from $\L''$, since the $R$ term contributes only at tree level. When working with $\L$ there are three more diagrams with two-point insertions (which can be on external legs)
\tikzfeynmanset{  insertion@@/.style args={[#1]#2}{
    /tikz/decoration={
      markings,
      mark=at position #2 with {
        \tikzfeynmanset{insertion/.cd,#1}
        \filldraw[draw=black] (-0.075,-0.075) rectangle ++(0.15,0.15);
      },
    },
    /tikz/postaction={
      /tikz/decorate=true,
    }  
}}
\begin{equation}
	\begin{tikzpicture}[baseline=-0.08cm]
    \begin{feynman}
      \vertex (a);
      \vertex [right=1cm of a] (b);
      \vertex [right=1cm of b] (c);
      \vertex [right=1cm of c] (d);
      \diagram* {
        a [crossed dot] -- [insertion={0.5}] b  -- [fermion] c -- [fermion] d [crossed dot],
        (b) -- [photon, half left] (c)
      };
    \end{feynman}
    \end{tikzpicture} \quad + \quad 
\begin{tikzpicture}[baseline=-0.08cm]
  \begin{feynman}
    \vertex (a);
    \vertex [right=1cm of a] (b);
    \vertex [right=1cm of b] (c);
    \vertex [right=1cm of c] (d);
    \diagram* {
            a [crossed dot] -- [fermion] b  -- [insertion={0.5}] c -- [fermion] d [crossed dot],
      (b) -- [photon, half left] (c)
    };
  \end{feynman}
\end{tikzpicture} \quad + \quad 
\begin{tikzpicture}[baseline=-0.08cm]
  \begin{feynman}
    \vertex (a);
    \vertex [right=1cm of a] (b);
    \vertex [right=1cm of b] (c);
    \vertex [right=1cm of c] (d);
    \diagram* {
            a [crossed dot] -- [fermion] b  -- [fermion] c -- [insertion={0.5}] d [crossed dot],
      (b) -- [photon, half left] (c)
    };
  \end{feynman}
\end{tikzpicture} \;.
\end{equation}
After adding up all diagrams, we find agreement for the $\psi$ propagator at one loop. When assuming the redundant operator to be generated through renormalization, so $R \sim \mathcal O(\text{1-loop})$, the two-loop propagator will automatically agree, since it is calculated from interactions with $R = 0$ (so vertices between $\L$ and $\L''$ agree). This verifies that up to the two-loop order, the $\psi$ propagator does not depend on wether one uses $\L$ or $\L''$. We performed the same check for the $\psi \bar\psi A$ vertex.

\subsection{Gauge-variant sub-divergences}
\label{sec:ClassIIbNuisance}

In the case of non-abelian gauge theories, we encounter class-IIb nuisance operators, which are not gauge invariant but only constrained by BRST symmetry. As an illustration, we use single-flavor QCD augmented by dimension-five operators~\cite{Misiak:1994zw}
\begin{align}
	\label{eq:QCDLagrangian}
	\L[\psi,J] &= -\frac{1}{4} G_{\mu\nu}^A G^{A\mu\nu} + \bar \psi \, (i \slashed D-m) \psi  + L \, \bar \psi \,\sigma_{\mu \nu} G^{\mu \nu} \psi \nn
		&\quad + R_1 \, \bar \psi \, (i \slashed{D} - m)^2 \psi  + R_2 \left( \bar \psi \, \slashed Q (i \slashed{D} - m) \psi - \bar \psi (i \overleftarrow{\slashed D} + m)  \slashed Q \psi \right) \nn
		&\quad + \L_\mathrm{gf} + \L_\mathrm{gh} \nn
		&\quad + \bar J \psi + \bar \psi J + J_\mu^A Q^{A\mu} \, ,
\end{align}
where $D_\mu = \p_\mu + i g G_\mu$, $G_\mu = t^A G_\mu^A$, $G_{\mu\nu} = t^A G^A_{\mu\nu}$, and the $SU(N_c)$ gauge field is split into background and quantum fields
\begin{equation}
	G_\mu^A = B_\mu^A + Q_\mu^A \, .
\end{equation}
The gauge-fixing term in the background-field method is
\begin{equation}
	\L_\mathrm{gf} = - \frac{1}{2 \xi} (G^A)^2 \, , \quad G^A = \p_\mu Q^\mu - g f^{ABC} B^B_\mu Q^{C\mu}
\end{equation}
and $\L_\mathrm{gh}$ is the ghost Lagrangian. The form of the single gauge-variant nuisance operator with coefficient $R_2$ is restricted by BRST and $CP$ symmetry, hermiticity, as well as by background-gauge invariance.
Both redundant operators can be removed by a field redefinition, which modifies the source terms according to
\begin{align}
	\L'[\psi,J] &= -\frac{1}{4} G_{\mu\nu}^A G^{A\mu\nu} + \bar \psi \, (i \slashed D-m) \psi  + L \, \bar \psi \,\sigma_{\mu \nu} G^{\mu \nu} \psi \nn
		&\quad + \L_\mathrm{gf} + \L_\mathrm{gh} \nn
		&\quad + \bar J (1-R_2 \slashed Q) \psi + \bar \psi(1-R_2 \slashed Q) J + R_1 \bar J J + \O(\text{dim-6}) \, .
\end{align}
In contrast to the abelian case, the sources couple non-linearly to the new fields, hence after the field redefinition the original Green's functions correspond to Green's functions of composite operators. Since any interpolating field with the correct quantum numbers can be used in the LSZ formula, this field redefinition shows that both the gauge-invariant and gauge-variant nuisance operators only lead to effects that are compensated by external-leg corrections, hence they do not affect the $S$-matrix.

We compute the off-shell renormalization of this EFT at two loops in the modified minimal-subtraction (\msbar{}) scheme. The complete results for the renormalization constants and RGEs up to two loops are given in App.~\ref{sec:QCDDipoleResults}. We will use it in the next section as a test case of different variants of the IR rearrangement. In particular, we will show that redundant operators do not only leave the $S$-matrix invariant, but that there is also a way to obtain the correct RGEs for the coefficients of physical operators from an off-shell calculation without considering the insertion of redundant class-II operators.


\section{Expansion of loop integrands and infrared rearrangement}
\label{sec:IRrearrangement}

Although EFTs contain a tower of higher-dimension operators, they are renormalizable order by order in the power counting, which is most transparent when using dimensional regularization. As in ordinary renormalizable theories, the counterterms for local EFTs are polynomials in masses and momenta, and can be determined recursively at each loop order after the cancellation of non-local sub-divergences of lower loop orders. This cancellation can be achieved globally at the level of Green's functions by computing separately the diagrams of lower loop order with counterterm insertions, or at the level of individual Feynman diagrams by making use of the $R$-operation. In Sect.~\ref{sec:Roperation}, we briefly discuss these methods as well as the Taylor expansion of integrands before integration, which introduces spurious IR divergences. In Sects.~\ref{sec:DummyMass} and~\ref{sec:IRdimreg}, we explain two different approaches to treat these IR divergences.

\subsection[Renormalization, $R$-operation, and expansion of integrands]{\boldmath Renormalization, $R$-operation, and expansion of integrands}
\label{sec:Roperation}

In the approach of \textit{global renormalization}, counterterms are identified with terms in the Lagrangian and constructed at the level of Green's function. Counterterm graphs, which cancel sub-divergences, are most conveniently obtained through shifts in the couplings $C \mapsto C + \delta C$ and multiplication with wave-function renormalization factors, which are performed after the perturbative expansion. The perturbative expansion is thus carried out in terms of the bare parameters. The counterterms that need to be explicitly inserted into counterterm diagrams are extracted from lower-loop Green's functions of both background and quantum fields. While the identification with terms in the Lagrangian provides a strong consistency check across different correlators, it can be cumbersome to perform these steps, in particular because the counterterms to sub-graphs with quantum fields involve class-IIb operators. The need to construct a potentially large number of gauge-variant operators (see, e.g., Ref.~\cite{Cirigliano:2020msr}) and to determine their counterterms is in contrast to the original motivation of the background-field method.

These complications are avoided in the $\bar R$-operation~\cite{Collins:1984xc,Smirnov:1985yck,Chetyrkin:1984xa,Herzog:2017bjx}, which can be employed to automatically subtract sub-divergences from individual graphs.\footnote{$\bar{R}$ only subtracts subdivergences, giving the local divergence. $R$ without the bar also adds the overall local counterterm, leading to a finite value.} It does so by identifying superficially divergent subgraphs and adding countergraphs. The countergraphs are obtained from the original graph by contracting the subgraph into a point and insertion of a counterterm, which is simply minus the divergence of $\bar R$ applied to the subgraph itself.

At the two-loop order there are two generic topologies of Feynman diagrams contributing to 1PI Green's functions: the sunset and the figure-eight topologies with three and two possible sub-divergences, respectively. Schematically, the action of $\bar R$ is
\begin{align}
	\label{eq:Rbarappl}
	\bar R \,\, \sunsetdiag &= \sunsetdiag + \ctdiagOneAlt + \ctdiagTwo + \ctdiagThreeAlt \, , \nn
	\bar R \,\, \infdiag &= \infdiag + \ctdiagOne + \ctdiagTwo \, ,
\end{align}
where the graphs can have any number of external lines (which are not drawn) and arbitrary particles in the loops (all of which are drawn as solid lines). The crosses denote insertions of counterterms into the contracted original graphs, labeled by the propagator chain $i$ with $i =1,2,3$. These chains contain loop momenta $k_1$, $k_2$, and $k_3 = k_1+k_2$, respectively. While the sum of one-loop countergraphs to all two-loop diagrams reproduces the sum of one-loop counterterm diagrams (including wave-function renormalization) in a global renormalization, the $\bar R$-operation splits up the counterterm contributions in such a way that each two-loop diagram together with its countergraphs gives a local divergence. In the $\bar R$-operation, all counterterms including evanescent operators as well as class-II operators, are automatically generated from the divergent subgraphs, and inserted back into countergraphs. The $\bar R$-operation is therefore simpler to use than global renormalization. The drawback is that there is no direct consistency check on counterterms.

We are interested in the UV divergent parts of Eq. (\ref{eq:Rbarappl}), which is a polynomial in masses and external momenta of a degree limited by the superficial degree of divergence of the graph. Therefore, we can apply a Taylor expansion directly at the level of the Feynman integrands, denoted by an operator  $T$, without changing the overall UV poles,
\begin{equation}
	\label{eq:TaylorExpansionOfIntegrands}
	\bar R \,\, \sunsetdiag = T \bar R \; \sunsetdiag \; - \text{(spurious IR poles)} + \text{finite} \;. 
\end{equation}
Since we consider off-shell 1PI Green's functions, the original diagram has no IR singularities. The expansion of the integrands renders all integrals scaleless, so we can directly set to zero any integral which is not overall logarithmically divergent.\footnote{In the auxiliary mass method this step affects the deformed theory, as explained in Sect. \ref{sec:theorydeformation}. } The expansion turns the original non-local finite parts of the subdivergence-subtracted graph into spurious IR divergences, while leaving the overall UV divergence unaltered. The original non-local sub-divergences of the two-loop graph become local as well and are both UV and IR divergent. However, non-local divergences are subtracted by $\bar R$ before the Taylor expansion. The same cancellation happens after expansion, i.e., $\bar R$ cancels terms that are at the same time UV and IR divergent and leaves only terms that are either UV or IR divergent. In order to determine the UV counterterms, we need to drop the IR divergences. In practice, the separation of UV and IR divergences is non-trivial if both are dimensionally regulated. This problem is solved by so-called IR rearrangement, which can be performed in many different ways, see, e.g., Ref.~\cite{Lang:2020nnl}.

\subsection{Auxiliary mass as IR regulator}
\label{sec:DummyMass}

\subsubsection{Separation of UV and IR singularities}

Our first method for extracting UV poles is along the lines of Refs.~\cite{Stockinger:2023ndm,Born:2024mgz}. It is based on the introduction of an auxiliary dummy mass $m$ as an IR regulator~\cite{Chetyrkin:1997fm}, which is inserted in all propagator denominators after Taylor expansion and after dropping integrals that are not overall log-divergent. We call this operation $\hat m$. Application of $\hat m$ leaves the UV poles unchanged, as they stem from logarithmically divergent integrals, but it regulates the IR divergences and replaces them by logarithms of $m$. Putting everything together, we have 
\begin{align}
	\label{eq:methodoverview}
	\bar R \,\, \sunsetdiag &= \hat m \, T \, \bar R \; \sunsetdiag \; - \text{(logs of $m$)} + \text{finite} \nn
		&= \hat m \, T \, \sunsetdiag + \hat m \, T \, \ctdiagOneAlt +  \hat m \, T \,\ctdiagTwo +  \hat m \, T \, \ctdiagThreeAlt \; - \text{(logs of $m$)} + \text{finite} \, . 
\end{align}
It is crucial that the exact same $\hat m$ and $T$ operators be used across all terms in Eq. (\ref{eq:methodoverview}), especially since different natural choices for $\hat m$ exist. E.g., for fermionic propagators one could define
\begin{equation}
	\hat m: \;\frac{i}{\slashed k} \mapsto \frac{i}{\slashed k - m} = \frac{i (\slashed k + m)}{k^2 - m^2} \quad\quad\quad \text{or instead} \quad\quad\quad \hat m: \frac{i}{\slashed k} = \frac{i \slashed k}{k^2} \mapsto \frac{i \slashed k}{k^2-m^2} \, . 
\end{equation}
We choose the latter definition of $\hat m$, but both are equally valid, as long as they are used consistently across all contributions.

A sublety of the method concerns cancellations of loop momenta $k^2$. Generally terms $k^2/k^2$, which can appear after Taylor expansion, should be simplified to speed up the calculation. But cancelling before application of $\hat m$ gives a different result than cancelling afterwards, with the difference being terms proportional to the dummy mass,
\begin{equation}
	\frac{k^2}{k^2} - \frac{k^2}{k^2-m^2} = - \frac{m^2}{k^2-m^2} \, .
\end{equation}
In other words, cancellation does not commute with $\hat m$. Again, different natural choices for cancellation prescriptions exist. Here, we apply $\hat m$ directly after Taylor expansion, without any prior cancellations, which is a simple prescription that ensures a consistent definition of $\hat m$ across all contributions. Cancellations are then carried out against denominators containing $m$.

Finally we end up with two-loop vacuum integrals with a single scale $m$. Due to the absence of external momenta in denominators, the tensor decomposition is simple. The result is thus written in terms of scalar integrals of the form
\begin{equation}
	I(n_1,n_2,n_3) = \int \frac{d^Dk_1}{(2\pi)^D} \frac{d^Dk_2}{(2\pi)^D} \frac{1}{(k_1^2-m^2)^{n_1}(k_2^2-m^2)^{n_2}(k_3^2-m^2)^{n_3}} \, ,
\end{equation}
which can be reduced via recursion relations to the well-known case $I(1,1,1)$~\cite{Chetyrkin:1997fm}.

\subsubsection{\boldmath Theory deformation due to $\hat m$}
\label{sec:theorydeformation}
Application of $\hat m \, T $ changes individual terms in Eq.~\eqref{eq:methodoverview} but leaves the overall UV divergence of the sum invariant. The individual terms change because $\hat m$ introduces a dummy mass. At the level of the Lagrangian these terms correspond to a theory deformation $\Delta \L_m$ by additional counterterms proportional to powers of $m$. Such counterterms can violate gauge or even BRST symmetry. When working in global renormalization with a dummy mass, these counterterms need to be constructed and determined at lower loop levels, see, e.g., Ref.~\cite{Gambino:2003zm}.

In the $\bar R$-operation, $\Delta \L_m$ terms are again obtained and inserted automatically, which renders explicit construction and determination unnecessary. In the end, Eq.~\eqref{eq:methodoverview} guarantees that there is no effect on the UV-divergence of the subdivergence-subtracted two-loop graph $\bar R \, G$. The result found with the dummy-mass method corresponds to the true result for the UV divergence of $\bar R \, G$, which is independent of $\Delta \L_m$. 

As mentioned in Sect. \ref{sec:Roperation} we drop power-divergent two-loop integrals from our calculation (while retaining power-divergent one-loop integrals which lead to $m^0$ terms). Since after Taylor expansion the only available scale is $m$, power-divergent integrals would contribute only to $m$-dependent two-loop counterterms. Therefore, this step drops $\Delta \L_m$ at the two-loop level, while leaving our physical theory of interest invariant.   

\subsubsection{\boldmath Scheme definitions and the $R$-operation}

The $R$-operation must be adapted to the specific definitions of the employed scheme. In our work, this means that $R$ has to account for (A) finite renormalizations and (B) for the definition of the complete operator basis, including evanescent operators.

We discuss issue (A) first. In \msbar{}, $\bar R$ applied to a two-loop graph subtracts the sub-divergences by inserting the divergent part of the subgraphs into countergraphs. Since we are not using \msbar{} in the LEFT (see Sect.~\ref{sec:LEFT}), we need to modify $\bar R$ to account for finite renormalizations. Let us define $R$ as the \msbar{} operator and $R_f$ as the version in our scheme with divergent and finite parts of renormalization constants, denoted as $\delta C = \delta C_\mathrm{div} + \delta C_\mathrm{fin}$. Putting loop orders in superscripts, the renormalized value of a two-loop diagram in our scheme is
\begin{align}
	\label{eq:finreninR}
	R_f \; \sunsetdiagnonums & = \sunsetdiagnonums  + \ctdiag \; \delta C^{(1)} + \ldots + \overallct \, \Big(\delta C^{(2)} + (\delta C^{(1)})^2\Big) \nn
		& = \sunsetdiagnonums + \ctdiag \; \Big( \delta C_\mathrm{div}^{(1)} +  \delta C_\mathrm{fin}^{(1)} \Big) + \ldots + \overallct \, \Big(\delta C^{(2)} + (C_\mathrm{div}^{(1)} + C_\mathrm{fin}^{(1)})^2 \Big) \nn
		& = \bar R \; \sunsetdiagnonums + \ctdiag \;  \delta C_\mathrm{fin}^{(1)} + \ldots + \overallct \, \Big( \delta C_\mathrm{div}^{(2)} + C_\mathrm{div}^{(1)} C_\mathrm{fin}^{(1)} \Big) + \text{finite} \, .
\end{align}
In the last equality we have written the result in terms of the \msbar{} $\bar R$ operator, which suggests a simple way of dealing with finite renormalizations in context of the $R$-operation: to find the result in our scheme, we apply the \msbar{} operator $\bar R$, which is easier to implement and produces the first term. In the second term, only the divergent part of the one-loop diagram contributes, since it multiplies a finite counterterm $\delta C_\mathrm{fin}^{(1)}$. The diagram can be replaced with the negative of the respective tree-level diagram, with $\delta C_\mathrm{div}^{(1)}$ inserted. Thus, we add the effects of finite renormalizations separately, as these effects can be obtained from the one-loop counterterms. We perform this last step globally at the level of the Green's functions. 

Concerning issue (B), we point out that while in the $R$-operation the counterterms are constructed automatically, they still need to match the definition of the evanescent operators. Specifically, if a first evanescent scheme with a physical operator $\O$ and an evanescent operator $\E$ is related to a second scheme by
\begin{equation}
	\O = \O' \, , \quad \E = \E' + \varepsilon \O' \, ,
\end{equation}
an \msbar{} subtraction in the first scheme
\begin{equation}
	\frac{a}{\varepsilon} \O + \frac{b}{\varepsilon} \E = \left( \frac{a}{\varepsilon} + b\right) \O' + \frac{b}{\varepsilon} \E'
\end{equation}
does not correspond to \msbar{} in the second scheme. Since in the $R$-operation one usually does not explicitly map the counterterms to operators, one has to be careful to apply the subtractions in agreement with the operator definitions. In the LEFT, this becomes relevant at dimension six, as we will discuss in an upcoming publication~\cite{Naterop:2025lzc}.

\subsection{Dimensional regularization of IR singularities}
\label{sec:IRdimreg}

\subsubsection{Separation of UV and IR singularities}
\label{sec:UVIRseparation}

The main goal of the IR rearrangement is to disentangle UV from IR singularities in Eq.~\eqref{eq:TaylorExpansionOfIntegrands}, i.e., in integrals obtained by a Taylor expansion of the integrands before integration. In pure dimensional regularization in $D = 4 - 2 \varepsilon$ dimensions, both singularities are treated by the same regulator $\varepsilon$, hence the difficulty consists in distinguishing $\varepsilon_\mathrm{UV}$ from $\varepsilon_\mathrm{IR}$. The method described in Sect.~\ref{sec:DummyMass} circumvents the complication by introducing by hand the auxiliary mass as an alternative IR regulator. This changes the theory and in particular the UV sub-divergences at lower loop levels, however without affecting the overall UV divergences. In the following, we present a new variant of the IR rearrangement that does not modify the UV structure at all, but rather achieves the IR rearrangement by separating the UV-divergent from IR-divergent contributions in Eq.~\eqref{eq:TaylorExpansionOfIntegrands} and thereby allows us to distinguish $\varepsilon_\mathrm{UV}$ from $\varepsilon_\mathrm{IR}$. We find this approach particularly useful in combination with global renormalization and the background-field method. Parts of our method have been presented in App.~A of Ref.~\cite{Jenkins:2023rtg}. As many other existing variants, it is based on the partial-fraction relation called {\em tadpole decomposition}, which for scalar propagators reads~\cite{Misiak:1994zw,Chetyrkin:1997fm}:
\begin{align}
	\label{eq:TadpoleDecompositionBoson}
	D_M(k+p) &= D_m(k) - D_m(k) \left(  D_M^{-1}(k+p) - D_m^{-1}(k) \right) D_M(k+p) \nn
		&= D_m(k) - i D_m(k) \left( M^2 - m^2  - p^2 - 2 k \cdot p \right) D_M(k+p) \, ,
\end{align}
where $m$ is an auxiliary dummy mass, $k$ denotes a generic loop momentum and $p$ an external momentum, and where
\begin{equation}
	D_M(p) = \frac{i}{p^2 - M^2} \, .
\end{equation}
In the case of fermionic propagators with generic non-Hermitian and non-diagonal mass matrices in flavor space, the decomposition reads~\cite{Naterop:2023dek}
\begin{equation}
	\label{eq:TadpoleDecompositionFermion}
	S_\psi(k+p) = S_m(k) - i S_m(k) \left( M_\psi P_L  + M_\psi^\dagger P_R - m  - \slashed p \right) S_\psi(k+p) \, .
\end{equation}
where
\begin{equation}
	S_\psi(p) = i \left( \slashed p - M_\psi P_L  - M_\psi^\dagger P_R  \right)^{-1} \, , \quad
	S_m(p) = i \left( \slashed p - m  \right)^{-1} = \frac{i ( \slashed p + m)}{p^2 - m^2} \, .
\end{equation}
For a two-loop diagram with loop momenta $k_1$, $k_2$, and $k_3 = k_1 + k_2$, we also use the following {\em disentanglement identities}~\cite{Jenkins:2023rtg}
\begin{align}
	\label{eq:DisentanglementIdentities}
	S_\psi(k_3+p) &= S_m(k_1) - i S_m(k_1) \left( M_\psi P_L  + M_\psi^\dagger P_R - m  - \slashed p - \slashed k_2 \right) S_\psi(k_3+p) \, , \nn
	S_\psi(k_3+p) &= S_m(k_2) - i S_m(k_2) \left( M_\psi P_L  + M_\psi^\dagger P_R - m  - \slashed p - \slashed k_1 \right) S_\psi(k_3+p) \, , \nn
	S_\psi(k_1+p) &= S_m(-k_2) - i S_m(-k_2) \left( M_\psi P_L  + M_\psi^\dagger P_R - m  - \slashed p - \slashed k_3 \right) S_\psi(k_1+p) \, .
\end{align}

Green's functions at two-loop accuracy are obtained from the sum of two-loop diagrams, one-loop diagrams with an insertion of a one-loop counterterm (or a one-loop wave-function renormalization factor), as well as tree-level diagrams with an insertion of a two-loop counterterm or two one-loop counterterms (or wave-function factors). At the level of amplitudes, we denote this by
\begin{equation}
	i \A^\mathrm{NNLO} = i \A^{2\text{-loop}} + i \A^{1\text{-loop,ct(1)}} + i \A^\text{ct(2)} \, .
\end{equation}
In order to extract the RGEs, we are interested in the $1/\varepsilon$ divergence of the two-loop counterterms as well as the one-loop counterterms in case of finite renormalizations, see Sect.~\ref{sec:LEFT}. In a global renormalization, the counterterm is determined by calculating the divergences of the sum of all two-loop diagrams and all the one-loop diagrams with counterterm insertion. While these two contributions are separately non-local, the sum of the two needs to be a polynomial in masses and momenta.

To start the discussion of our IR rearrangement procedure, we consider the application of tadpole decompositions~\eqref{eq:TadpoleDecompositionBoson} and~\eqref{eq:TadpoleDecompositionFermion} and disentanglement identities~\eqref{eq:DisentanglementIdentities} before any Taylor expansion, i.e., the decomposition consists of the application of exact identities. Following the algorithm described in App.~A of Ref.~\cite{Jenkins:2023rtg}, these relations allow us to split each two-loop diagram into a sum of massive divergent two-loop tadpole (vacuum) integrals, products of two one-loop diagrams, and UV-finite terms, which can be discarded. Next, we apply the tadpole decomposition both in the product of two one-loop diagrams as well as the one-loop counterterm diagrams, schematically:
\begin{align}
	\A^\mathrm{NNLO} &= \A^{2\text{-loop}}_\text{tadpole} + \A^{1\text{-loop}}_A \times \A^{1\text{-loop}}_B + \A^{1\text{-loop}}_C \times \A^\text{ct(1)} + \A^\text{ct(2)} + \text{finite} \nn
		&= \A^{2\text{-loop}}_\text{tadpole} + \left( \A^{1\text{-loop}}_{A,\text{tadpole}} +  \A^{1\text{-loop}}_{A,\text{finite}} \right) \times \left( \A^{1\text{-loop}}_{B,\text{tadpole}} +  \A^{1\text{-loop}}_{B,\text{finite}} \right) \nn
			&\quad + \left( \A^{1\text{-loop}}_{C,\text{tadpole}} +  \A^{1\text{-loop}}_{C,\text{finite}} \right) \times \A^\text{ct(1)} + \A^\text{ct(2)} + \text{finite} \, .
\end{align}
We split off the UV divergences as follows:
\begin{align}
	\A^{2\text{-loop}}_\text{tadpole} &= \frac{N_\text{tadpole}^{(2,2)}}{\varepsilon^2} + \frac{N_\text{tadpole}^{(2,1)}}{\varepsilon} + \text{finite} \, , \nn
	\A^{1\text{-loop}}_\text{tadpole} &= \frac{N_\text{tadpole}^{(1,1)}}{\varepsilon} + \A^{1\text{-loop}}_\text{tadpole,finite} \, , \nn
	\A^\text{ct(2)} &= \frac{N_\text{ct}^{(2,2)}}{\varepsilon^2} + \frac{N_\text{ct}^{(2,1)}}{\varepsilon} \, , \nn
	\A^\text{ct(1)} &= \frac{N_\text{ct}^{(1,1)}}{\varepsilon} \, ,
\end{align}
which results in
\begin{align}
	N_\text{ct}^{(2,2)} &= - N_\text{tadpole}^{(2,2)} - N_{A,\text{tadpole}}^{(1,1)} \times N_{B,\text{tadpole}}^{(1,1)} - N_{C,\text{tadpole}}^{(1,1)} \times N_\text{ct}^{(1,1)} \, , \nn
	N_\text{ct}^{(2,1)} &= - N_\text{pure tadpoles}^{(2,1)} - N_\text{rest}^{(2,1)} \, ,
\end{align}
where
\begin{align}
	N_\text{pure tadpoles}^{(2,1)} &= N_\text{tadpole}^{(2,1)} + N_{A,\text{tadpole}}^{(1,1)} \times \A_{B,\text{tadpole,finite}}^{1\text{-loop}} + N_{B,\text{tadpole}}^{(1,1)} \times \A_{A,\text{tadpole,finite}}^{1\text{-loop}} \nn
		&\quad + N_\text{ct}^{(1,1)} \times \A_{C,\text{tadpole,finite}}^{1\text{-loop}} \, , \nn
	N_\text{rest}^{(2,1)} &= N_{A,\text{tadpole}}^{(1,1)} \times \A_{B,\text{finite}}^{1\text{-loop}} + N_{B,\text{tadpole}}^{(1,1)} \times \A_{A,\text{finite}}^{1\text{-loop}} + N_\text{ct}^{(1,1)} \times \A_{C,\text{finite}}^{1\text{-loop}} \, .
\end{align}
The counterterm $N_\text{ct}^{(2,1)}$ is a polynomial in the physical masses and momenta. $N_\text{pure tadpoles}^{(2,1)}$ is already manifestly a polynomial in the physical masses and momenta, but it also depends on the auxiliary dummy-mass parameter $m$ introduced by the tadpole decomposition. It follows that also $N_\text{rest}^{(2,1)}$ is a polynomial in the physical masses and momenta, i.e., non-local (logarithmic) contributions need to cancel between the different terms in $N_\text{rest}^{(2,1)}$. Since $N_\text{rest}^{(2,1)}$ is given by finite integrals, it must be a rational function of $m$ and the same then applies to $N_\text{pure tadpoles}^{(2,1)}$, because the sum of $N_\text{pure tadpoles}^{(2,1)}$ and $N_\text{rest}^{(2,1)}$ is independent of $m$.

An important simplification of the calculation results from the fact that the integrands of $N_\text{rest}^{(2,1)}$ can be expanded in all the physical masses and momenta before integration: this is allowed provided that no overall IR divergences are generated. Without expansion, IR divergences are manifestly absent, or naturally regulated by the masses and momenta. Terms resulting in IR divergences manifest themselves as singularities in the physical scales. While individual integrals do contain such singularities, they cancel in the sum of all integrals in $N_\text{rest}^{(2,1)}$, which is a polynomial in the physical scales. Likewise, upon expansion the IR divergences have to cancel in $N_\text{rest}^{(2,1)}$ and expanding the integrands before performing the loop integral does not change the result for $N_\text{rest}^{(2,1)}$. This works even if IR divergences in the expanded $N_\text{rest}^{(2,1)}$ are regulated dimensionally, provided that no $\varepsilon_\mathrm{UV}$-dependence is erroneously turned into an $\varepsilon_\mathrm{IR}$-dependence. The correct results are obtained if in $N_\text{rest}^{(2,1)}$ one first performs the Laurent expansion around $\varepsilon_\mathrm{UV} = 0$ and evaluates the residues $N^{(1,1)}$ of the $1/\varepsilon_\mathrm{UV}$ poles. The contraction of these residues with the finite one-loop integrals can be performed before applying the Taylor expansion. Since this expression is finite, one can remove the regulator, i.e., the residues of the $1/\varepsilon_\mathrm{UV}$ poles and all possible contractions are evaluated for $D=4$. In a final step, one continues the finite integrals again to $D = 4 - 2 \varepsilon_\mathrm{IR}$ dimensions, applies a Taylor expansion of the integrands and performs all the remaining algebra, including the tensor decomposition of the UV-finite one-loop integrals, with the regulator $\varepsilon_\mathrm{IR}$.

Since the tadpole decompositions in this procedure are identities, they commute with the Taylor expansion. The same results for the UV-divergences are obtained if the Taylor expansion is performed in the very beginning, which simplifies further the entire computation. As a result, the IR rearrangement takes the following form.
\begin{enumerate}
	\item expand all integrands of two-loop diagrams and one-loop counterterm diagrams in masses and external momenta
	\item keep only overall log-divergent integrals
	\item apply the two-loop tadpole decomposition and disentanglement identities to split the scaleless two-loop integrals into massive divergent two-loop tadpoles, products of one-loop integrals, and UV-finite integrals
	\item drop UV-finite integrals
	\item apply the one-loop tadpole decomposition to the product of one-loop diagrams as well as the one-loop counterterm diagrams
	\item drop UV-finite terms in the products of one-loop integrals
	\item evaluate pure (two-loop and one-loop) tadpole contributions as usual
	\item evaluate the residue of $1/\varepsilon_\mathrm{UV}$ poles in $N_\text{rest}^{(2,1)}$, performing all possible contractions and sending $\varepsilon_\mathrm{UV}\to0$ in the residue
	\item evaluate the final UV-finite one-loop integrals in $D=4-2\varepsilon_\mathrm{IR}$ dimensions, including any potentially remaining tensor decomposition or integration-by-part reduction
\end{enumerate}
We checked that this procedure leads to the same results as the one of Sect.~\ref{sec:DummyMass} for the theories discussed in Sects.~\ref{sec:EOMOperators} and~\ref{sec:ClassIIbNuisance}. In contrast to the method with auxiliary-mass IR regulator, no additional $m$-dependent counterterms are generated. The method also does not rely on a specific uniform choice of momentum routing, hence it is straightforward to treat two-loop diagrams and one-loop counterterm diagrams separately in a global renormalization. This can lead to a reduction of the computational cost compared to the local $R$-operation. We expect that the procedure can be generalized to higher loop orders, which we leave for future work.

\subsubsection{Omission of class-II operators}

Using the described IR rearrangement within a global renormalization instead of the local $R$-operation seems to reintroduce the need to consider class-II operators, in particular gauge-variant class-IIb nuisance operators appearing as sub-divergences. However, it turns out that the entire renormalization of the physical sector can be performed without considering nuisance operators.

As discussed in Sect.~\ref{sec:NuisanceOperators}, redundant operators do not influence the calculation of $S$-matrix elements: neglecting them would lead to incorrect 1PI Green's functions, but the discussed field redefinitions make clear that the modifications affect also the external-leg corrections in such a way that they drop out upon amputation in the LSZ formula. Hence, it is common practice to neglect redundant operators in the on-shell calculation of $S$-matrix elements. While it is possible to renormalize the physical sector directly on shell, the computation of $S$-matrix elements is in general more involved than the computation of the 1PI effective action.

In contrast to $S$-matrix elements, the situation is different in the calculation of the 1PI effective action. In this case, the divergences arising in off-shell 1PI Green's function need to be mapped to the complete basis of the theory, including on-shell-redundant operators. In the background-field method, this includes only correlators of background fields, which are gauge invariant. However, this is only true for the highest considered loop level: in order to cancel sub-divergences, one has to consider class-IIb operators as explained in Sect.~\ref{sec:NuisanceGeneralities}. If one neglects the counterterms to redundant operators, one does not obtain the correct off-shell 1PI Green's function and the result will contain uncanceled sub-divergences, which are non-local. A priori, it is not clear how these spurious non-local divergences could be unambiguously mapped to the operator basis in a way that does not affect the physical sector.

The solution is provided by the procedure described in Sect.~\ref{sec:UVIRseparation}. The Taylor expansion turns the result manifestly into a polynomial in masses and momenta, which can always be mapped onto the operator basis. The IR rearrangement splits UV divergences from spurious IR singularities. In the sum of two-loop diagrams and one-loop counterterm diagrams, the dependence on the dummy mass $m$ drops out. If the full operator basis is included, also all spurious IR singularities vanish in the UV-divergent part. IR singularities only remain in the UV-finite part, which is split off by the IR rearrangement. If we do not include the full operator basis, but neglect the insertion of redundant operators, the result for the UV-divergent part of the 1PI effective action contains spurious IR singularities, i.e., terms proportional to $1/(\varepsilon_\mathrm{UV} \times \varepsilon_\mathrm{IR})$. However, due to the prior Taylor expansion, the result is still a polynomial in masses and momenta that can be mapped onto the operator basis. Doing so, the spurious IR singularities reside in the unphysical sector of the theory, i.e., they affect the counterterms to redundant operators as well as wave-function renormalization factors, which are not physical. The counterterms of the coefficients of physical operators are not affected by the uncanceled IR singularities and they are unaltered from the correct result of the full calculation. If it were different, ignoring redundant operators would also affect the $S$-matrix, which can be composed from vertices of the 1PI effective action inserted into tree-level topologies.

We have verified this procedure in the two-loop renormalization of the theories described in Sects.~\ref{sec:EOMOperators} and~\ref{sec:ClassIIbNuisance}. Indeed, using the IR rearrangement of Sect.~\ref{sec:UVIRseparation} we find that neglecting the insertion of redundant operators only modifies the counterterms of the redundant operators and wave-function renormalization factors, while leaving the physical counterterms unchanged, which also fully agree with the results obtained with the $R$-operation.

The described method provides an efficient procedure to renormalize the physical sector in an off-shell calculation. Using the background-field method, no gauge-variant class-IIb operators need to be considered.


\section{Renormalization-group equations of the LEFT at two loops}
\label{sec:LEFT}

In this section, we discuss the two-loop RGEs of the LEFT up to dimension five in the power counting. Our calculation is carried out in the scheme defined in Ref.~\cite{Naterop:2023dek}, which we first briefly review in Sect.~\ref{sec:LEFTOpsScheme}. In Sect.~\ref{sec:Calculation}, we mention some details of the calculation. We discuss a few aspects of our results in Sect.~\ref{sec:Results}, whereas the entire two-loop RGEs are provided in App.~\ref{sec:LEFTRGE}.

\subsection{Operator basis and scheme definition}
\label{sec:LEFTOpsScheme}

The Lagrangian for the LEFT is 
\begin{equation}
	\label{eq:LEFTLagrangian}
	\L_\mathrm{LEFT} = \L_\mathrm{QCD+QED} + \L_{\nu} + \sum_{d\ge5} \sum_i \lwc{i}{(d)} \O_i^{(d)} \, ,
\end{equation}
with the leading-order terms
\begin{align}
	\label{eq:qcdqed}
	\L_{\rm QCD + QED} &= - \frac14 G_{\mu \nu}^A G^{A \mu \nu} -\frac14 F_{\mu \nu} F^{\mu\nu} + \theta_{\rm QCD} \frac{g^2}{32 \pi^2} G_{\mu \nu}^A \widetilde G^{A \mu \nu} +  \theta_{\rm QED} \frac{e^2}{32 \pi^2} F_{\mu \nu} \widetilde F^{\mu \nu} \nn
		&\quad + \sum_{\psi=u,d,e}\overline \psi \left( i \slashed D - M_\psi P_L - M_\psi^\dagger P_R \right) \psi
\end{align}
and covariant derivative $D_\mu = \p_\mu + i g T^A G_\mu^A + i e Q A_\mu$. The LEFT power counting is dictated by the expansion parameter $p/v$ or $m/v$, where $v$ denotes the electroweak scale, $p$ an external momentum, and $m$ a mass of the degrees of freedom in the theory. The effective operators up to dimension five are given in Tab.~\ref{tab:basisphysandeom} (physical and on-shell redundant operators) and Tab.~\ref{tab:basisevanescent} (evanescent operators). We work with the basis defined in Ref.~\cite{Naterop:2023dek}. Note that in contrast to the sample theories in Sects.~\ref{sec:EOMOperators} and~\ref{sec:ClassIIbNuisance}, the redundant operators are not written in manifest EOM form, i.e., in our basis the actual class-IIa operators consist of linear combinations of physical, on-shell redundant, and evanescent operators. 

\begin{table}[t]
	\capstart
	\begin{adjustbox}{width=0.9\textwidth,center}
		\begin{minipage}[t]{3cm}
			\renewcommand{\arraystretch}{1.51}
			\small
			\begin{align*}
			\begin{array}[t]{c|c}
			\multicolumn{2}{c}{\boldsymbol{(\nu \nu) X+\hc}} \\
			\hline
			\O_{\nu \gamma} & (\nu_{Lp}^T C   \bar\sigma^{\mu \nu}  \nu_{Lr})  F_{\mu \nu}  \\
			\end{array}
			\end{align*}
			\end{minipage}			
		\begin{minipage}[t]{3cm}
			\renewcommand{\arraystretch}{1.51}
			\small
			\begin{align*}
			\begin{array}[t]{c|c}
			\multicolumn{2}{c}{\boldsymbol{(\overline L R ) X+\hc}} \\
			\hline
			\O_{e \gamma} & \bar e_{Lp}   \bar\sigma^{\mu \nu} e_{Rr}\, F_{\mu \nu}  \\
			\O_{u \gamma} & \bar u_{Lp}   \bar\sigma^{\mu \nu}  u_{Rr}\, F_{\mu \nu}   \\
			\O_{d \gamma} & \bar d_{Lp}  \bar\sigma^{\mu \nu} d_{Rr}\, F_{\mu \nu}  \\
			\O_{u G} & \bar u_{Lp}   \bar\sigma^{\mu \nu}  T^A u_{Rr}\,  G_{\mu \nu}^A  \\
			\O_{d G} & \bar d_{Lp}   \bar\sigma^{\mu \nu} T^A d_{Rr}\,  G_{\mu \nu}^A \\
			\end{array}
			\end{align*}
		\end{minipage}
		\begin{minipage}[t]{3cm}
			\renewcommand{\arraystretch}{1.51}
			\small
			\begin{align*}
			\begin{array}[t]{c|c}
			\multicolumn{2}{c}{\boldsymbol{(\nu\nu)D^2 + \hc}} \\
			\hline
			\O_{\nu D}^{(5)} & \nu_{Lp}^T C (i\bar{\slashed \p})^2 \nu_{Lr} \\
			\end{array}
			\end{align*}
			\end{minipage}
			\begin{minipage}[t]{3cm}
				\renewcommand{\arraystretch}{1.51}
				\small
				\begin{align*}
				\begin{array}[t]{c|c}
				\multicolumn{2}{c}{\boldsymbol{(\overline L R)D^2 + \hc}} \\
				\hline
				\O_{eD}^{(5)} & \bar e_{Lp} (i\bar{\slashed D})^2 e_{Rr} \\
				\O_{uD}^{(5)} & \bar u_{Lp} (i\bar{\slashed D})^2 u_{Rr} \\
				\O_{dD}^{(5)} & \bar d_{Lp} (i\bar{\slashed D})^2 d_{Rr} \\
				\end{array}
				\end{align*}
				\end{minipage}							
		\end{adjustbox}
	\caption{Physical operators in the LEFT at dimension five (columns 1 and 2), as well as on-shell redundant operators, which can be removed via field redefinitions (columns 3 and 4), reproduced from Ref.~\cite{Naterop:2023dek}.}
	\label{tab:basisphysandeom}
\end{table}

\begin{table}[t]
	\capstart
	\begin{adjustbox}{width=0.9\textwidth,center}
		\begin{minipage}[t]{3cm}
			\renewcommand{\arraystretch}{1.51}
			\small
			\begin{align*}
			\begin{array}[t]{c|c}
			\multicolumn{2}{c}{\boldsymbol{(\nu \nu)D + \hc}} \\
			\hline
			\E_{\nu D}^{} & \nu_{Lp}^T C (i\hat{\slashed \p}) \nu_{Lr} \\
			\end{array}
			\end{align*}
		\end{minipage}
		\begin{minipage}[t]{3cm}
			\renewcommand{\arraystretch}{1.51}
			\small
			\begin{align*}
			\begin{array}[t]{c|c}
			\multicolumn{2}{c}{\boldsymbol{(\overline L R)D + \hc}} \\
			\hline
			\E_{e D}^{} & \bar e_{Lp} (i\hat{\slashed D}) e_{Rr} \\
			\E_{u D}^{} & \bar u_{Lp} (i\hat{\slashed D}) u_{Rr} \\
			\E_{d D}^{} & \bar d_{Lp} (i\hat{\slashed D}) d_{Rr} \\
			\end{array}
			\end{align*}
		\end{minipage}
		\begin{minipage}[t]{3cm}
			\renewcommand{\arraystretch}{1.51}
			\small
			\begin{align*}
			\begin{array}[t]{c|c}
			\multicolumn{2}{c}{\boldsymbol{X^2}} \\
			\hline
			\E_{\gamma} & F_{\mu\nu} F^{\mu\nu} - \roverline{F_{\mu\nu} F^{\mu\nu}} \\
			\E_{G} & G_{\mu\nu}^A G^{A\mu\nu} - \roverline{G_{\mu\nu}^A G^{A\mu\nu}} \\
			\E_{\gamma'} & \hat F_{\mu\nu} \hat F^{\mu\nu} \\
			\E_{G'} & \hat G_{\mu\nu}^A \hat G^{A\mu\nu} \\
			\end{array}
			\end{align*}
		\end{minipage}			
		\begin{minipage}[t]{3cm}
			\renewcommand{\arraystretch}{1.51}
			\small
			\begin{align*}
			\begin{array}[t]{c|c}
			\multicolumn{2}{c}{\boldsymbol{X^2}} \\
			\hline
			\E_{\gamma} & F_{\mu\nu} F^{\mu\nu} - \roverline{F_{\mu\nu} F^{\mu\nu}} \\
			\E_{G} & G_{\mu\nu}^A G^{A\mu\nu} - \roverline{G_{\mu\nu}^A G^{A\mu\nu}} \\
			\E_{\gamma'} & \hat F_{\mu\nu} \hat F^{\mu\nu} \\
			\E_{G'} & \hat G_{\mu\nu}^A \hat G^{A\mu\nu} \\
			\end{array}
			\end{align*}
		\end{minipage}			
	\end{adjustbox}
	\begin{adjustbox}{width=0.8\textwidth,center}
		\begin{minipage}[t]{3cm}
			\renewcommand{\arraystretch}{1.51}
			\small
			\begin{align*}
			\begin{array}[t]{c|c}
			\multicolumn{2}{c}{\boldsymbol{(\overline L L)D^2}} \\
			\hline
			\E_{\nu D}^{L} & \bar \nu_{Lp} [(i\hat{\slashed \p}) , (i\bar{\slashed \p})] \nu_{Lr} \\
			\E_{e D}^{L} & \bar e_{Lp} [(i\hat{\slashed D}) , (i\bar{\slashed D})] e_{Lr} \\
			\E_{u D}^{L} & \bar u_{Lp} [(i\hat{\slashed D}) , (i\bar{\slashed D})] u_{Lr} \\
			\E_{d D}^{L} & \bar d_{Lp} [(i\hat{\slashed D}) , (i\bar{\slashed D})] d_{Lr} \\
			\end{array}
			\end{align*}
		\end{minipage}			
		\begin{minipage}[t]{3cm}
			\renewcommand{\arraystretch}{1.51}
			\small
			\begin{align*}
			\begin{array}[t]{c|c}
			\multicolumn{2}{c}{\boldsymbol{(\overline R R)D^2}} \\
			\hline
			\E_{e D}^{R} & \bar e_{Rp} [(i\hat{\slashed D}) , (i\bar{\slashed D})] e_{Rr} \\
			\E_{u D}^{R} & \bar u_{Rp} [(i\hat{\slashed D}) , (i\bar{\slashed D})] u_{Rr} \\
			\E_{d D}^{R} & \bar d_{Rp} [(i\hat{\slashed D}) , (i\bar{\slashed D})] d_{Rr} \\
			\end{array}
			\end{align*}
		\end{minipage}		
		\begin{minipage}[t]{3cm}
			\renewcommand{\arraystretch}{1.51}
			\small
			\begin{align*}
			\begin{array}[t]{c|c}
			\multicolumn{2}{c}{\boldsymbol{(\nu \nu)D^2 + \hc}} \\
			\hline
			\E_{\nu D}^{RL} & \nu_{Lp}^T C (i\hat{\slashed \p})(i\hat{\slashed \p}) \nu_{Lr} \\
			\end{array}
			\end{align*}
		\end{minipage}
		\begin{minipage}[t]{3cm}
			\renewcommand{\arraystretch}{1.51}
			\small
			\begin{align*}
			\begin{array}[t]{c|c}
			\multicolumn{2}{c}{\boldsymbol{(\overline L R)D^2 + \hc}} \\
			\hline
			\E_{e D}^{LR} & \bar e_{Lp} (i\hat{\slashed D})(i\hat{\slashed D}) e_{Rr} \\
			\E_{u D}^{LR} & \bar u_{Lp} (i\hat{\slashed D})(i\hat{\slashed D}) u_{Rr} \\
			\E_{d D}^{LR} & \bar d_{Lp} (i\hat{\slashed D})(i\hat{\slashed D}) d_{Rr} \\
			\end{array}
			\end{align*}
		\end{minipage}			
	\end{adjustbox}
	\begin{adjustbox}{width=0.97\textwidth,center}
		\begin{minipage}[t]{3cm}
			\renewcommand{\arraystretch}{1.51}
			\small
			\begin{align*}
			\begin{array}[t]{c|c}
			\multicolumn{2}{c}{\boldsymbol{(\overline L L)X}} \\
			\hline
			\E_{\nu\gamma}^{L} & (\bar\nu_{Lp} i\hat\gamma_\mu\bar\gamma_\nu \nu_{Lr})  F^{\mu\nu}  \\
			\E_{e\gamma}^{L} & (\bar e_{Lp} i\hat\gamma_\mu\bar\gamma_\nu e_{Lr})  F^{\mu\nu}  \\
			\E_{u\gamma}^{L} & (\bar u_{Lp} i\hat\gamma_\mu\bar\gamma_\nu u_{Lr})  F^{\mu\nu}  \\
			\E_{d\gamma}^{L} & (\bar d_{Lp} i\hat\gamma_\mu\bar\gamma_\nu d_{Lr})  F^{\mu\nu}  \\
			\E_{uG}^{L} & (\bar u_{Lp} i\hat\gamma_\mu\bar\gamma_\nu T^A u_{Lr})  G^{A\mu\nu}  \\
			\E_{dG}^{L} & (\bar d_{Lp} i\hat\gamma_\mu\bar\gamma_\nu T^A d_{Lr})  G^{A\mu\nu}  \\
			\end{array}
			\end{align*}
		\end{minipage}
		\begin{minipage}[t]{3cm}
			\renewcommand{\arraystretch}{1.51}
			\small
			\begin{align*}
			\begin{array}[t]{c|c}
			\multicolumn{2}{c}{\boldsymbol{(\overline R R)X}} \\
			\hline
			\E_{e\gamma}^{R} & (\bar e_{Rp} i\hat\gamma_\mu\bar\gamma_\nu e_{Rr})  F^{\mu\nu}  \\
			\E_{u\gamma}^{R} & (\bar u_{Rp} i\hat\gamma_\mu\bar\gamma_\nu u_{Rr})  F^{\mu\nu}  \\
			\E_{d\gamma}^{R} & (\bar d_{Rp} i\hat\gamma_\mu\bar\gamma_\nu d_{Rr})  F^{\mu\nu}  \\
			\E_{uG}^{R} & (\bar u_{Rp} i\hat\gamma_\mu\bar\gamma_\nu T^A u_{Rr})  G^{A\mu\nu}  \\
			\E_{dG}^{R} & (\bar d_{Rp} i\hat\gamma_\mu\bar\gamma_\nu T^A d_{Rr})  G^{A\mu\nu}  \\
			\end{array}
			\end{align*}
		\end{minipage}			
		\begin{minipage}[t]{3cm}
			\renewcommand{\arraystretch}{1.51}
			\small
			\begin{align*}
			\begin{array}[t]{c|c}
			\multicolumn{2}{c}{\boldsymbol{(\nu\nu)X + \hc}} \\
			\hline
			\E_{\nu\gamma}^{RL} & (\nu_{Lp}^T C \hat\sigma_{\mu\nu} \nu_{Lr})  F^{\mu\nu}  \\
			\end{array}
			\end{align*}
		\end{minipage}	
		\begin{minipage}[t]{3cm}
			\renewcommand{\arraystretch}{1.51}
			\small
			\begin{align*}
			\begin{array}[t]{c|c}
			\multicolumn{2}{c}{\boldsymbol{(\overline L R)X + \hc}} \\
			\hline
			\E_{e\gamma}^{LR} & (\bar e_{Lp} \hat\sigma_{\mu\nu} e_{Rr})  F^{\mu\nu}  \\
			\E_{u\gamma}^{LR} & (\bar u_{Lp} \hat\sigma_{\mu\nu} u_{Rr})  F^{\mu\nu}  \\
			\E_{d\gamma}^{LR} & (\bar d_{Lp} \hat\sigma_{\mu\nu} d_{Rr})  F^{\mu\nu}  \\
			\E_{uG}^{LR} & (\bar u_{Lp} \hat\sigma_{\mu\nu} T^A u_{Rr})  G^{A\mu\nu}  \\
			\E_{dG}^{LR} & (\bar d_{Lp} \hat\sigma_{\mu\nu} T^A d_{Rr})  G^{A\mu\nu}  \\
			\end{array}
			\end{align*}
		\end{minipage}
	\end{adjustbox}
	\caption{ Evanescent operators in the LEFT at dimension four (first row) and five (second and third row), reproduced from Ref.~\cite{Naterop:2023dek}.}
	\label{tab:basisevanescent}
\end{table}

We split $G_{\mu}^A$ into background and quantum gluons and implement the standard background-field gauge fixing~\cite{Abbott:1980hw,Abbott:1983zw}. The field-strength tensors and dual field-strength tensors for photons and gluons are
\begin{align}
    F_{\mu\nu} &= \p_\mu A_\nu - \p_\nu A_\mu \, , \quad\widetilde F^{\mu\nu} = \frac{1}{2} \epsilon^{\mu\nu\lambda\sigma} F_{\lambda\sigma} \, , \nn
    G_{\mu\nu}^A & = \p_\mu G_\nu^A - \p_\nu G_\mu^A - g f^{ABC} G_\mu^B G_\nu^C\, , \quad \widetilde G^{A\mu\nu} = \frac{1}{2} \epsilon^{\mu\nu\lambda\sigma} G^A_{\lambda\sigma} \, ,
\end{align}
with the Levi-Civita sign convention $\epsilon_{0123} = +1$. For the definition of $\L_\nu$ and the remaining conventions used we refer to Ref.~\cite{Naterop:2023dek}.
We work in $D = 4-2\varepsilon$ dimensions and we use the HV scheme for $\gamma_5$~\cite{tHooft:1972tcz}.
Our operator basis defines effective operators from the physical basis over four (instead of $4-2\varepsilon$) space-time dimensions, for instance the electron dipole operator is
\begin{equation}
	\L = \lwc{e\gamma}{}[][pr] \, \bar e_{Lp} \, \bar \sigma^{\mu\nu} \, e_{Rr} \, F_{\mu\nu} + \text{h.c.} + \ldots \, ,
\end{equation}
with flavor indices $p,r$ and the bar restricting the Lorentz indices to $\mu,\nu\in\{0,1,2,3\}$. Using a $D$-dimensional $\sigma^{\mu\nu}$ instead corresponds to an evanescent shift, which is a scheme change under which the one-loop RGE is invariant. The two-loop RGE however is not: in general, it depends on the scheme, including the $\gamma_5$ prescription, the operator definition, and the renormalization conditions. In particular, it is affected by finite counterterms. In our scheme, the parameters $X_i$ of the theory are renormalized at one loop according to
\begin{equation}
	X_i = \mu^{n_i \varepsilon} ( X_i^r(\mu) + X_i^\mathrm{ct} ),   \quad \quad  X_i^\text{ct} = X_{i,\text{div}} + X_{i,\text{fin}}^\chi+ X_{i,\text{fin}}^\text{ev} \, ,
\end{equation}
where $X_{i,\text{div}}$ cancel all one-loop divergences and in addition we define finite renormalizations to take care of two effects: the compensation of evanescent-operator insertions through $X_{i,\text{fin}}^\text{ev}$, as well as the restoration of chiral symmetry in the spurion sense through  $X_{i,\text{fin}}^\chi$. Both the divergent and the finite counterterms $X_{i,\text{div}}$, $X_{i,\text{fin}}^\text{ev}$, and $X_{i,\text{fin}}^\text{chi}$ are provided explicitly in Ref.~\cite{Naterop:2023dek}. The finite one-loop renormalizations affect both the finite parts of one-loop Green's functions as well as the two-loop RGEs, as will be discussed below. We treat gauge couplings, masses, and Wilson coefficients on an equal footing. In general, the counterterms $X_i^\mathrm{ct}$ are expanded in loops and powers of $\varepsilon$,
\begin{align}
	X_i^\mathrm{ct} &= \sum_{l=1}^\infty \sum_{n=0}^l \frac{1}{\varepsilon^n} \frac{1}{(16\pi^2)^l} X_i^{(l,n)}(\{L_j^r(\mu)\},\{K_k^r(\mu)\}) \, ,
\end{align}
where $L_i$ and $K_i$ denote physical and evanescent Wilson coefficients. Note that we do not perform finite renormalizations in the evanescent sector, i.e., $K_i^{(l,0)} = 0$.

\subsection{Loop calculation}
\label{sec:Calculation}

\begin{figure}[t]
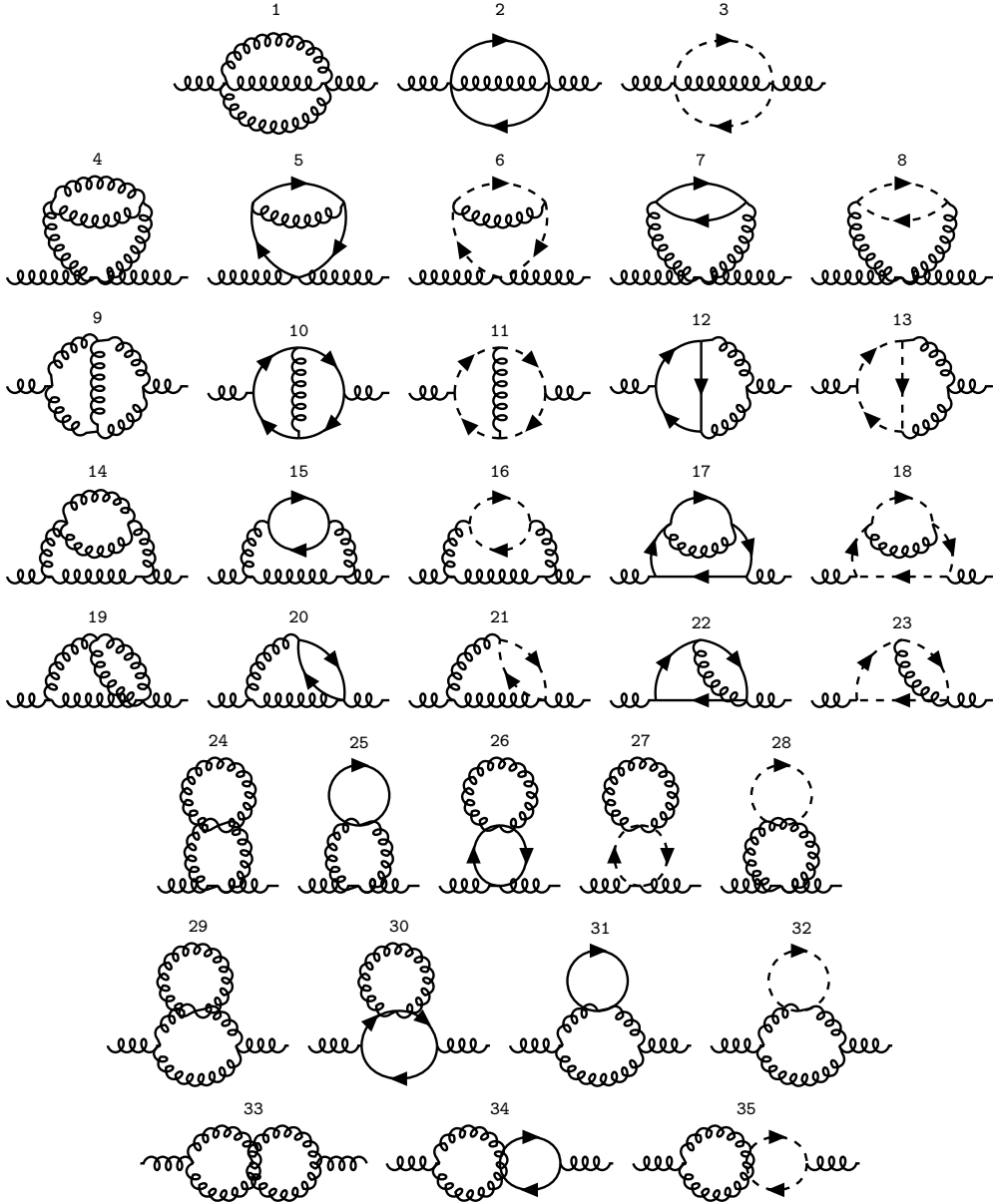

	\centering
	\gluondiagDA \gluondiagDB \gluondiagDC
	\\ 
	\gluondiagEA \gluondiagEB \gluondiagEC \gluondiagED \gluondiagEE
	\\ \vspace{0.2cm}
	\gluondiagFA \gluondiagFB \gluondiagFC \gluondiagFD \gluondiagFE
	\\ \vspace{0.2cm}
	\gluondiagGA \gluondiagGB \gluondiagGC \gluondiagGD \gluondiagGE	
	\\ \vspace{0.2cm}
	\gluondiagHA \gluondiagHB \gluondiagHC \gluondiagHD \gluondiagHE
	\\ \vspace{0.2cm}
	\gluondiagAA \gluondiagAB \gluondiagAC \gluondiagAD\gluondiagAE
	\\ \vspace{0.2cm}
	\gluondiagBA \gluondiagBB \gluondiagBC \gluondiagBD
	\\ 
	\gluondiagCA \gluondiagCB \gluondiagCC
	\caption{The two-loop diagrams for the gluon propagator in dimension-five LEFT. Diagrams obtained from those by exchange of external legs or reflection of a fermion loop are not displayed. There exist also diagrams with different fermions in loops, and diagrams with internal photons, totaling to 83 diagrams (see Tab. \ref{tab:amountdiags}). Some diagrams vanish trivially, e.g., due to the EFT power counting. }
	\label{fig:gluondiags}
\end{figure}

We compute the renormalization of the LEFT diagrammatically. At dimension five and two loops, this involves the computation of 1491 Feynman diagrams. A breakdown into different Green's functions is given in Tab.~\ref{tab:amountdiags}. As an example, we show in Fig.~\ref{fig:gluondiags} explicitly the topologies for the gluon propagator. 

\begin{table}[t]
	\centering
	\begin{tabular}{l|l|l|l|l|l|l|l|l|l|l|l|l|l}
		$AA$ & $GG$ & $ee$ & $uu$ & $dd$ & $eeA$ & $uuA$ & $ddA$ & $uuG$ & $ddG$ & $GGG$ & $AA\,\delta\zeta$ & $GG\,\delta\zeta$ & \textbf{total} \\ \hline
		21 & 83 & 6  & 41 & 41 & 17  & 80  & 80  & 207 & 207 &  487  & 79      & 142      & \textbf{1491}
	\end{tabular}
	\caption{Number of diagrams for the different Green's functions computed for the two-loop renormalization of the LEFT at dimension five. The labels mean $e =$ lepton, $u =$ up-type quark, $d=$ down-type quark, $A=$ photon, $G=$ gluon, and $\delta\zeta$ = dummy field used in the calculation of the $\theta$ terms. The $GGG$ Green's function is not independently needed at dimension five but serves as a cross-check. }
	\label{tab:amountdiags}
\end{table}

We use \texttt{QGRAF} \cite{Nogueira:1991ex} for diagram generation and our own routines for the derivation and application of Feynman rules. Small expressions are processed in \texttt{Mathematica}. For critical steps with large expressions we make heavy use of \texttt{FORM}~\cite{Vermaseren:2000nd,Ruijl:2017dtg} and \texttt{Symbolica}. The slowest step in the calculation is Dirac-algebra simplification: in the HV scheme, due to the presence of $\gamma_5$ in Dirac chains and traces, the required dimensional split results in a large number of terms. A comparable calculation using NDR is found to be two to three orders of magnitudes faster. In the HV scheme, the calculation is considerably faster when one uses Hermitian mass matrices instead of generic masses: due to the absence of projectors from mass insertions, the bulk of the Dirac simplification can then be carried out without a dimensional split, thus speeding up the calculation by one to two orders of magnitudes.

The RGEs are derived from the UV countertems. In the presence of finite renormalizations, the relation at two loops is~\cite{Naterop:2023dek}
\begin{equation}
	\label{eq:rgeoverview}
	\dot L = 
		4 L_i^{(2,1)} - \sum_j 2 L_j^{(1,0)} \frac{\p L_i^{(1,1)}}{\p L_j} - \sum_j 2 L_j^{(1,1)} \frac{\p L_i^{(1,0)}}{\p L_j} - \sum_j  2 K_j^{(1,1)} \frac{\p L_i^{(1,0)}}{\p K_j} \, ,
\end{equation}
where for \msbar{} only the first term contributes. In our scheme with finite shifts the remaining terms also contribute: they cancel (A)~the dependence on the coefficients $K_i$ of evanescent operators and (B)~chiral-symmetry-breaking terms, both of which can be present in the first term of Eq.~\eqref{eq:rgeoverview}. The cancellation (A) was not checked as it would require evanescent insertions into two-loop diagrams.\footnote{This is expected to be a computationally expensive task due to the large number of evanescent operators.} Instead, we assign a loop counting to evanescent operators and assume their coefficients to be one-loop generated, hence we neglect all dependence on the coefficients $K_i$ in two-loop terms and the two-loop RGEs are only affected by evanescent-operator insertions into one-loop diagrams. The cancellation (B) can be verified explicitly and provides a useful check of the consistency across loop orders: indeed, all symmetry-breaking terms are absent in our results for the RGEs.

The cancellation of sub-divergences is done automatically with the $R$-operation in \msbar{}; finite renormalizations were accounted for separately and globally, see Sect.~\ref{sec:Roperation}. As a result, counterterms and RGEs were obtained in \msbar{} as well as in our scheme with finite shifts.

The complete calculation is based on two independent implementations of the $R$-operation (one in \texttt{FORM}, the other in \texttt{Symbolica}) described in Sect.~\ref{sec:DummyMass} and a cross-check of the resulting counterterms and RGEs. In addition, we performed partial cross-checks based on a global renormalization and the IR rearrangement presented in Sect.~\ref{sec:IRdimreg}. The entire calculation was done with generic QED and QCD quantum gauge parameters $\xi_\gamma$ and $\xi_g$. Gauge-parameter independence of the RGEs of physical operators provides another useful check on the calculation.

\subsection{Results}
\label{sec:Results}

In App.~\ref{sec:LEFTRGE}, we provide the results for the LEFT RGEs in our chirally symmetric scheme at dimension five, i.e., including all effects from a single insertion of a dimension-five operator suppressed by $1/v$ in the power counting. The results are given for the physical sector of the theory after having performed the field redefinitions that remove redundant operators.

Our chirally symmetric HV scheme has a more complicated counterterm structure than \msbar{}. However, it is interesting to note that Green's functions and RGEs are more compact due to the absence of evanescent and chiral-symmetry-breaking terms. We emphasize that in the LEFT, which is a vector-like gauge theory, these finite renormalizations are a scheme choice and one could consistently calculate as well in the \msbar{} scheme. While \msbar{} is the natural choice for pure QED and QCD, the presence of chiral structures in higher-dimension operators makes our chirally symmetric scheme much more appealing: without the finite symmetry-restoring counterterms, one needs to take into account chiral-symmetry-breaking terms in matching contributions, in RGEs, as well as matrix elements. Overall, these effects have to cancel each other in relations between observables. In our scheme, chiral-symmetry-breaking terms are manifestly absent not only in relations between observables, but also in intermediate steps, such as matching contributions or RGEs.

As an example, omitting from Eq.~\eqref{eq:eRGE} the contribution of quarks, the two-loop RGE for the electric charge in \msbar{} is (see Eq.~\eqref{eq:RGENotation} for the notation)
\begin{equation}
	\label{eq:exampleRGEMS}
	[\dot{e}]_2^\text{\msbar{}}  = 
		4 \; n_e e^5 \q_e^4 + 48 \, e^4 \q_e^3 \Big( \< L_{e\gamma} M_e^\dag \> + \< L_{e\gamma}^\dag M_e \> \Big) - \frac{224}{3} e^4 \q_e^3 \Big( \< L_{e\gamma} M_e \> + \< L_{e\gamma}^\dag M_e^\dag \> \Big) \, ,
\end{equation}
where $\< \cdot \>$ denotes the trace in flavor space, $n_e$ is the number of charged lepton flavors, and $\q_e = -1$ the electron charge. In contrast, our scheme with finite shifts leads to the simpler and chirally symmetric result
\begin{equation}
	\label{eq:exampleRGE}
	[\dot{e}]_2  =
		4 \; n_e e^5 \q_e^4 - 80 e^4 \q_e^3 \Big(  \< L_{e\gamma} M_e \> + \< L_{e\gamma}^\dag M_e^\dag \>  \Big) \, ,
\end{equation}
which is free of spurion-symmetry-breaking terms of the form $L M^\dag$, $L^\dag M$, as expected. We note that the QED parts of Eqs.~\eqref{eq:exampleRGEMS} and~\eqref{eq:exampleRGE} agree, as they should. The reason is that this term can be extracted from a single-parameter calculation, for which the scheme dependence starts at three loops. 

In addition to the gauge couplings $e$ and $g$, we compute the RGEs for the QED and QCD $\theta$-parameters, using the method of Ref.~\cite{Georgi:1980cn}. The results are given in Eqs.~\eqref{eq:thetaQEDRGE} and~\eqref{eq:thetaQCDRGE}. Considering the linear combinations \\[-0.2cm]
\begin{equation}
	\tau_\mathrm{QED} = i \frac{4\pi}{e^2} + \frac{\theta_\mathrm{QED}}{2\pi} \, , \quad \tau_\mathrm{QCD} = i \frac{4\pi}{g^2} + \frac{\theta_\mathrm{QCD}}{2\pi} \, ,
\end{equation}
we find that our scheme respects holomorphy at two loops and dimension five in the case of $\tau_\mathrm{QED}$~\cite{Alonso:2014rga,Jenkins:2017dyc}, but we observe a violation of holomorphy in $\tau_\mathrm{QCD}$ proportional to $N_c$, leading to a dependence of the RGE for $\tau_\mathrm{QCD}$ on the anti-self-dual gluonic dipole-operator coefficients $L_{uG}^\dagger$ and $L_{dG}^\dagger$. However, we note that these are scheme-dependent statements, which are affected by finite renormalizations and the definition of physical and evanescent operators. We expect that one could perform an additional finite renormalization that restores holomorphy of $\tau_\mathrm{QCD}$ at dimension five.

We apply our chirally symmetric HV scheme throughout. In the case of pure QED or QCD, the finite renormalizations lead to unusual RGEs for the masses. In the limit of pure QED, the anomalous dimension of the electron mass in our scheme is given by
\begin{equation}
	[\dot M_e ]_2 = - e^4 \q_e^4 (4 n_e + 3) M_e \, ,
\end{equation}
which is chirally symmetric in the spurion sense. In contrast, the \msbar{} result in the HV scheme
\begin{equation}
	[\dot M_e ]_2^\text{\msbar{}} = -e^4 \q_e^4 (4 n_e + 3) M_e +  e^4 \q_e^4 \frac{32}{3} n_e M_e^\dagger
\end{equation}
contains a chiral-symmetry-breaking contribution proportional to $M_e^\dagger$. In the case of a Hermitian mass matrix, $m_e := M_e = M_e^\dagger$, the \msbar{} expression reduces to the well-known result
\begin{equation}
	[\dot m_e ]_2^\text{\msbar{}} = e^4 \q_e^4 \left( \frac{20}{3} n_e - 3 \right) m_e \, ,
\end{equation}
which agrees with the NDR scheme because no $\gamma_5$ appears in QED with a Hermitian mass matrix. Analogously, in the limit of pure QCD our scheme leads to the quark-mass anomalous dimension
\begin{equation}
	[\dot M_q ]_2 = - g^4 C_F \left( 3 (C_F + N_c) + 2 n_q \right) M_q \, ,
\end{equation}
where $n_q$ denotes the number of quark flavors, whereas in \msbar{}, the HV scheme produces an additional symmetry-breaking contribution,
\begin{equation}
	[\dot M_q ]_2^\text{\msbar{}} = - g^4 C_F \left( 3 (C_F + N_c) + 2 n_q \right) M_q  - \frac{8}{3} g^4 C_F \left( 11 N_c - 2 n_q \right) M_q^\dagger \, .
\end{equation}
For a Hermitian mass matrix $m_q := M_q = M_q^\dagger$, the \msbar{} result reduces to the known expression
\begin{equation}
	[\dot m_q ]_2^\text{\msbar{}} = - g^4 C_F \left( 3 C_F + \frac{97}{3}N_c - \frac{10}{3} n_q \right) m_q \, ,
\end{equation}
which again coincides with the result in the NDR scheme.

The complete results for the two-loop RGEs of the mass matrices are given in App.~\ref{sec:RGEsMasses}. We stress that these results are not unique: since we are working in a basis of generic non-diagonal and non-Hermitian mass matrices, there is always the freedom to perform further chiral rotations in flavor space, which changes the mass matrices, see Refs.~\cite{Jenkins:2017dyc,Dekens:2019ept,Naterop:2023dek}. Indeed, we find chirally symmetric and gauge-parameter-independent two-loop RGEs for the mass matrices only after performing such a chiral field redefinition at the two-loop level. Regarding the gauge-parameter independence, the same observation was already made in Refs.~\cite{Dekens:2019ept,Naterop:2023dek}. The necessary chiral field redefinitions turn out to be purely axial rotations, hence they affect the $\theta$-parameters at the three-loop level. Due to the freedom of performing chiral rotations, our results in App.~\ref{sec:RGEsMasses} should be understood as the definition of the field basis. As in Refs.~\cite{Dekens:2019ept,Naterop:2023dek}, we choose the gauge-parameter-dependent axial rotation to be proportional to $\xi_{\gamma}-1$ or $\xi_g-1$. The ambiguity in the basis choice drops out when one computes a matrix element: at this point, one has to apply a transformation to the mass basis, which renders the mass matrices real and diagonal.

The chiral rotation that brings us to the basis in which we report our results is suppressed by one factor of the LEFT power counting. Therefore, at dimension five it has no effect beyond the mass matrices. This will no longer be the case at dimension six~\cite{Naterop:2025}.


\section{Conclusions and outlook}
\label{sec:Conclusions}

In this article, we have presented the first part of a complete calculation of the two-loop RGEs of the LEFT in a HV scheme that respects chiral symmetry in the spurion sense, as previously established in Ref.~\cite{Naterop:2023dek}. The HV scheme is the only one that is proven to be consistent to all loop orders~\cite{Breitenlohner:1977hr}, but due to the dimensional split it comes with significant computational challenges.

Here, we have concentrated on several technical aspects of the calculation: we have elaborated on the advantages of the background-field method and the problem of gauge-variant nuisance operators in sub-divergences. We have discussed how the construction and identification of these class-IIb operators can be avoided, either by making use of the automatic subtraction in the $R$-operation, or by avoiding their insertion completely, using a modification of the IR rearrangement. In both cases, one can obtain the renormalization of the physical sector of the theory from an off-shell calculation of 1PI Green's functions.

We have subsequently presented the results for the complete LEFT two-loop RGEs up to dimension five in our scheme. They consist of the two-loop renormalization of the dipole operators, as well as their mixing into the fermion masses and the gauge couplings, including the $\theta$-parameters, which in the case of $\theta_\mathrm{QCD}$ is of phenomenological interest in the context of searches for a neutron electric dipole moment~\cite{Abel:2020pzs}. When using our results, it is important that all parts of the calculation be performed in the same scheme (or scheme changes be correctly taken into account), in particular matching calculations for the LEFT, either at the electroweak scale~\cite{Dekens:2019ept} or the hadronic scale~\cite{Mereghetti:2021nkt,Buhler:2023gsg,Crosas:2023anw}. Adapting these matching calculations to our chirally invariant HV scheme is work in progress~\cite{Crosas:2025}.

The techniques and methods presented here are not restricted to dimension five. The entire results for the LEFT RGEs at dimension six, which have a wider range of phenomenological applications, will be presented in forthcoming publications~\cite{Naterop:2025lzc,Naterop:2025}. They are part of a wider effort to establish a complete EFT framework for physics beyond the SM at next-to-leading-log accuracy.

	
	\section*{Acknowledgements}
	\addcontentsline{toc}{section}{\numberline{}Acknowledgements}

	We thank B.~Grinstein, A.~V.~Manohar, B. Ruijl, C.-H.~Shen, A.~Signer, D.~St\"ockinger, A.~E.~Thomsen, and M.~Zoller for useful discussions.
	Financial support by the Swiss National Science Foundation (Project No.~PCEFP2\_194272) is gratefully acknowledged.

	\clearpage

	
	\appendix


\section{\boldmath Counterterms and RGEs for $CP$-even dipole operators}
\label{sec:Counterterms}

Below we give the \msbar{} renormalization constants and RGEs for the $U(1)$ and $SU(3)$ EFT models discussed in Sects.~\ref{sec:EOMOperators} and~\ref{sec:ClassIIbNuisance}. We use the short-hand notation $\{ X \}_1 = X/(16\pi^2)$ and $\{ X \}_2 = X / (16\pi^2)^2$ for the one-loop and two-loop contributions, respectively.
In contrast to our full LEFT calculation, the results in this appendix are obtained in a scheme where the dipole operators are defined in $D$ dimensions and pure \msbar{} is employed. Since no $\gamma_5$ shows up in these toy EFTs, no dimensional split is needed and there is no difference between NDR and HV prescriptions.

\subsection[Single-flavor QED with $CP$-even dipole]{\boldmath Single-flavor QED with $CP$-even dipole}
\label{sec:QEDDipoleResults}

For the theory defined in Eq.~\eqref{eq:QEDLagrangian} we find the renormalization constants (before field redefinition)
\begin{align}
        \delta Z_A &= \biggl\{ 
            -\frac{4 e \q (e \q-12 L m)}{3 \varepsilon }
        \biggr\}_1 + \biggl\{
            \frac{16 e^3 \q^3 L m}{\varepsilon ^2} -\frac{2 e^3 \q^3 (e \q-24 L m)}{\varepsilon }
        \biggr\}_2 \, , \nn
        \delta Z_\psi &= \biggl\{ 
            \frac{e \q (6 L m -e  \q \xi)}{\varepsilon }
        \biggr\}_1 + \biggl\{
            \frac{e^3 \q^3 \left(e \q \xi ^2+4 L m (5 - 3 \xi )\right)}{2 \varepsilon ^2} + \frac{e^3 \q^3 (21 e \q-652 L m)}{12 \varepsilon }
        \biggr\}_2 \, , \nn
        \delta m &= \biggl\{ 
            -\frac{3 e \q m (e \q-4 L m)}{\varepsilon }
        \biggr\}_1 + \biggl\{
            \frac{e^3 \q^3 m (5 e \q+16 L m)}{2 \varepsilon ^2} + \frac{11 e^3 \q^3 m (e \q-112 L m)}{12 \varepsilon }
        \biggr\}_2 \, , \nn
        \delta e &= \biggl\{ 
            \frac{2 e^2 \q (e \q-12 L m)}{3 \varepsilon }
        \biggr\}_1 + \biggl\{
            \frac{2 e^4 \q^3 (e \q-36 L m)}{3 \varepsilon ^2}+\frac{e^4 \q^3 (e \q-24 L m)}{\varepsilon }
        \biggr\}_2 \, , \nn
        \delta L &= \biggl\{ 
            \frac{17 e^2 \q^2 L}{3 \varepsilon }
        \biggr\}_1 + \biggl\{
            \frac{119 e^4 \q^4 L}{6 \varepsilon ^2}-\frac{587 e^4 \q^4 L}{36 \varepsilon }
        \biggr\}_2 \, , \\
        \delta R &= \biggl\{ 
            -\frac{e \q (e  \q R \xi - 6 L)}{\varepsilon }
        \biggr\}_1 + \biggl\{
            \frac{e^3 \q^3 \left(e \q R \xi ^2 + L (38 - 6 \xi )\right)}{2 \varepsilon ^2} + 
            \frac{e^3 \q^3 (21 e \q R+4 L (3 \xi - 70))}{12 \varepsilon }
        \biggr\}_2 . \nonumber
\end{align}
The coefficient of the redundant operator $R$ only mixes into $R$ itself and does not affect the physical sector. The field redefinition sets $\delta R=0$ and leaves the remaining renormalization constants unchanged. The $l$-loop anomalous dimensions in \msbar \, are simply given by $2l$ times the $1/\varepsilon$ poles (using the notation $\dot{x} = \mu \frac{d}{d \mu} x$)
\begin{align}
        \dot{m} &= \biggl\{ 
            -6 e^2 \q^2 m + 24 e \q L m^2
        \biggr\}_1 + \biggl\{ 
            \frac{11}{3} e^4 \q^4 m - \frac{1232}{3} e^3 \q^3 L m^2
        \biggr\}_2 \, , \nn
        \dot{e} &= \biggl\{ 
            \frac{4 e^3 \q^2}{3}-16 e^2 \q L m
        \biggr\}_1 + \biggl\{ 
            4 e^5 \q^4-96 e^4 \q^3 L m
        \biggr\}_2 \, , \nn
        \dot{L} &= \biggl\{ 
            \frac{34}{3} e^2 \q^2 L
        \biggr\}_1 + \biggl\{ 
            -\frac{587}{9} e^4 \q^4 L
        \biggr\}_2 \, .
\end{align}

\subsection[Single-flavor QCD with $CP$-even dipole]{\boldmath Single-flavor QCD with $CP$-even dipole}
\label{sec:QCDDipoleResults}

Below, we list the \msbar{} renormalization constants and RGEs for the case of QCD with a $CP$-even $D$-dimensional dipole operator, as defined in Sect.~\ref{sec:ClassIIbNuisance}. We treat the theory using the background-field method.
{\small
\begin{align}
        \delta{Z_B} &= \biggl\{ 
            \frac{g (g (11 N_c - 2)+24 L m)}{3 \varepsilon }
        \biggr\}_1 \nn
        	&\quad + \biggl\{ 
            -\frac{4 g^3 L m \left(N_c^2 + 1\right)}{N_c \varepsilon ^2} + \frac{g^3\left(g \left(34 N_c^3 - 13 N_c^2 + 3\right)+24 L m \left(8 N_c^2 - 3\right)\right)}{6 N_c \varepsilon }
        \biggr\}_2   \, , \nn
        \delta{Z_Q} &= \biggl\{ 
            \frac{g (g (N_c (13 - 3 \xi )-4)+48 L m)}{6 \varepsilon }
        \biggr\}_1 \nn & \quad + \biggl\{ 
            \frac{ g^4 N_c (2\xi+3)(N_c(3\xi-13)+4)- 48 g^3 L m \left(N_c^2(2\xi+5) + 2\right)/N_c}{24 \varepsilon ^2} \nn
            &\qquad + \frac{g^3 \left(g \left(N_c^3(-2\xi^2-11\xi+59) - 28 N_c^2 + 8\right)+16 L m \left(25 N_c^2-12\right)\right)}{16 N_c \varepsilon }
        \biggr\}_2 \, , \nn
        \delta{Z_\psi} &= \biggl\{ 
            -\frac{g \left(N_c^2 - 1\right) (g \xi -6 (L-R_2) m)}{2 N_c \varepsilon }
        \biggr\}_1 \nn
        &\quad + \biggl\{ 
            \frac{g^3 \left(N_c^2 - 1\right) \left(g \xi  \left(N_c^2 (2 \xi + 3) - \xi \right)+L m \left(-(12 \xi  + 65) N_c^2 + 8 N_c + 12 (\xi  - 1)\right)\right)}{8 N_c^2 \varepsilon ^2}       \nn
            &\qquad + \frac{g^3 \left(N_c^2 - 1\right) R_2 m \left( (6 \xi  + 31) N_c^2 - 4 N_c - 6 \xi  - 9\right)}{4 N_c^2 \varepsilon ^2}       
            \nn & \qquad -\frac{g^3 \left(N_c^2 - 1\right) \left(3 g \left(\left(\xi ^2 + 8 \xi  + 22\right) N_c^2 - 4 N_c + 3\right)+L m \left(-559 N_c^2 + 280 N_c - 372\right) \right)}{48 N_c^2 \varepsilon  }
            \nn & \qquad +\frac{g^3 \left(N_c^2 - 1\right) R_2 m \left(-203 N_c^2 + 20 N_c + 9\right)}{24 N_c^2 \varepsilon  }
        \biggr\}_2  \, , \nn
        \delta{m} &= \biggl\{ 
            -\frac{3 g m \left(N_c^2 - 1\right) (g-4 L m)}{2 N_c \varepsilon }
        \biggr\}_1 \nn
        &\quad + \biggl\{ 
            \frac{g^3 m \left(N_c^2 - 1\right) \left(g \left(31N_c^2 - 4 N_c - 9\right)+8 L m \left(-23 N_c^2 + 8 N_c + 6\right)\right)}{8 N_c^2 \varepsilon ^2} \nn & \qquad + \frac{g^3 m \left(N_c^2-1\right) \left(g \left(-203 N_c^2+20 N_c+9\right)+8 L m \left(145 N_c^2-124 N_c+30\right)\right)}{48 N_c^2 \varepsilon }
        \biggr\}_2 \, , \nn
        \delta{g} &= \biggl\{ 
            - \frac{g^2 (g (11 N_c - 2)+24 L m)}{6 \varepsilon }
        \biggr\}_1 \nn & \quad + \biggl\{ 
            \frac{g^5 (2 - 11 N_c)^2+ 48 g^4 L m \left(12 N_c^2 - 2 N_c + 1\right)/N_c}{24\varepsilon ^2} \nn
            &\qquad + \frac{g^4 \left(g \left(-34 N_c^3 + 13 N_c^2 - 3\right)+24 L m \left(3 - 8 N_c^2\right)\right)}{12 N_c \varepsilon }
        \biggr\}_2 \, , \nn
        \delta{L} &= \biggl\{ 
            -\frac{g^2 L \left(8 N_c^2 - 2 N_c + 15\right)}{6 N_c \varepsilon }
        \biggr\}_1 \nn & \quad + \biggl\{ 
            \frac{g^4 L \left(80 N_c^4 - 36 N_c^3 + 194 N_c^2 - 40 N_c + 75\right)}{24 N_c^2 \varepsilon ^2} \nn
            &\qquad - \frac{g^4 L \left(161 N_c^4 - 164 N_c^3 + 1196 N_c^2 - 128 N_c + 459\right)}{144 N_c^2 \varepsilon }
        \biggr\}_2  \, , \nn
        \delta{R_1} &= \biggl\{ 
            \frac{g \left(N_c^2 - 1\right)(6 L - g R_1 \xi + 2 R_2 \xi)}{2 N_c \varepsilon }
        \biggr\}_1  + \biggl\{ 
            -\frac{g^3 L \left(N_c^2 - 1\right) \left((3 \xi  + 38) N_c^2 - 8 N_c - 6 (\xi  - 5)\right)}{8 N_c^2 \varepsilon ^2} 
            \nn & \quad + \frac{g^3 \xi (g R_1 - 2 R_2) \left(N_c^2-1\right) \left(N_c^2(2\xi+3) - \xi\right)}{8 N_c^2 \varepsilon^2 }
            \nn & \quad + \frac{g^3 L \left(N_c^2 - 1\right) \left((6 \xi  + 457) N_c^2 - 88 N_c - 12 (\xi  - 16)\right)}{48 N_c^2 \varepsilon }
            \nn & \quad - \frac{g^3 (g R_1 - 2 R_2) \left(N_c^2 - 1\right) \left((\xi^2+8 \xi  + 22) N_c^2 - 4 N_c + 3\right)}{16 N_c^2 \varepsilon }
        \biggr\}_2  \, , \nn
        \delta{R_2} &= \biggl\{ 
            \frac{g^2 ( 9 N_c L + R_2(4-22N_c))}{12 \varepsilon }
        \biggr\}_1 \nn & \quad + \biggl\{ 
            \frac{ g^4\left( 2 R_2(2-11N_c)^2 - 3 L (41N_c^2 - 8 N_c + 15) \right)}{48 \varepsilon ^2} \nn
            &\qquad + \frac{g^4 \left( L N_c(N_c^2(9\xi+83) - 20 N_c + 69) - 8 R_2(34N_c^3-13N_c^2 + 3) \right)}{96 N_c \varepsilon }
        \biggr\}_2 \, .
\end{align}}%
The expressions for the physical sector ($m$, $g$, $L$) are $\xi$-independent. For the RGEs we obtain
\begin{align}
        \dot{m} &= \biggl\{ 
            -\frac{3 g m \left(N_c^2 - 1\right) (g-4 L m)}{N_c}
        \biggr\}_1 \\
        	&\quad + \biggl\{ 
            \frac{g^3 m \left(1 - N_c^2\right) \left(g \left(203 N_c^2 - 20 N_c - 9\right)-8 L m \left(145 N_c^2 - 124 N_c + 30\right)\right)}{12 N_c^2}
        \biggr\}_2 \, , \nn
        \dot{g} &= \biggl\{ 
            -\frac{1}{3} g^2 (g (11 N_c - 2)+24 L m)
        \biggr\}_1  + \biggl\{ 
            -\frac{g^4 \left(g \left(34 N_c^3 - 13 N_c^2 + 3\right)+24 L m \left(8 N_c^2 - 3\right)\right)}{3 N_c}
        \biggr\}_2 \, , \nn
        \dot{L} &= \biggl\{ 
            -\frac{g^2 L \left(8 N_c^2 - 2 N_c + 15\right)}{3 N_c}
        \biggr\}_1  + \biggl\{ 
            -\frac{g^4 L \left(161 N_c^4 - 164 N_c^3 + 1196 N_c^2 - 128 N_c + 459\right)}{36 N_c^2}
        \biggr\}_2 \, . \nonumber
\end{align}
The result for $\dot L$ agrees with Ref.~\cite{Misiak:1994zw}, the remaining terms with $L$ are new to the best of our knowledge. For $N_c=3$ the expressions reduce to 
\begin{align}
        \dot{m} &= \biggl\{ 
            -8 g^2 m+ 32 \, g L m^2
        \biggr\}_1  + \biggl\{ 
            -\frac{1172 \, g^4 m}{9}+\frac{1712}{3} g^3 L m^2
        \biggr\}_2 \, , \nn
        \dot{g} &= \biggl\{ 
            -\frac{31 g^3}{3}-8 g^2 L m
        \biggr\}_1  + \biggl\{ 
            -\frac{268 g^5}{3}-184 \, g^4 L m
        \biggr\}_2 \, , \nn
        \dot{L} &= \biggl\{ 
            -9 g^2 L
        \biggr\}_1  + \biggl\{ 
            -\frac{1621}{27} g^4 L
        \biggr\}_2 \, .
\end{align}


\section{LEFT RGEs at dimension five in the HV scheme}
\label{sec:LEFTRGE}

In the following, we provide the results for the RGEs of the LEFT in our chirally symmetric HV scheme, at dimension five in the power counting, i.e., including all effects resulting from a single insertion of a dimension-five operator suppressed by $1/v$. Since neutrino interactions start at dimension five, at this order there is no dependence on the number of neutrino flavors $n_\nu$. We compute the complete set of RGEs, including the mixing into the QED and QCD theta parameters. We use the short-hand notation
\begin{equation}
	\label{eq:RGENotation}
	\dot X = \frac{d}{d\log\mu} X = \frac{1}{16\pi^2} [\dot X]_1 + \frac{1}{(16\pi^2)^2} [\dot X]_2
\end{equation}
and we only list the two-loop contribution $[\dot X]_2$ to the RGEs; the scheme-independent one-loop contribution $[\dot X]_1$ to the RGEs has been computed to dimension six in Ref.~\cite{Jenkins:2017dyc} (and independently confirmed in Ref.~\cite{Naterop:2023dek}). All RGE results are given in a compact matrix notation in flavor space. We define the running of the gauge couplings in pure QED plus QCD by
\begin{align}
	[\dot e]_\ell = - e \sum_{k=0}^{\ell-1} b^e_{\ell-1,k} \, e^{2(\ell-k)} g^{2k} \, , \quad [\dot g]_\ell = - g \sum_{k=0}^{\ell-1} b^g_{\ell-1,k} \, g^{2(\ell-k)} e^{2k} \, ,
\end{align}
where the coefficients of the $\beta$-functions up to two loops are
\begin{align}
	b^e_{0,0} &= -\frac{4}{3} \left( n_e \q_e^2 + N_c (n_u \q_u^2 + n_d \q_d^2 ) \right) \, , \nn
	b^g_{0,0} &= \frac{11}{3} N_c - \frac{2}{3} ( n_u + n_d ) \, , \nn
	b^e_{1,0} &= -4 \left( n_e \q_e^4 + N_c (n_u \q_u^4 + n_d \q_d^4) \right) \, , \; &
	b^e_{1,1} &= -4 N_c C_F \left( n_u \q_u^2 + n_d \q_d^2 \right) \, , \nn
	b^g_{1,0} &= \frac{34}{3} N_c^2 - 2 C_F ( n_u + n_d ) - \frac{10}{3} N_c ( n_u + n_d ) \, , \; &
	b^g_{1,1} &= -2 \left( n_u \q_u^2 + n_d \q_d^2 \right) \, ,
\end{align}
with $C_F = (N_c^2-1)/(2N_c)$ the coefficient of the quadratic Casimir operator in the fundamental representation of $SU(N_c)$. In the following, we use $C_F$ to write the results in a compact way, but we do not claim them to be valid for gauge groups different from $SU(N_c)$. In particular, we do not distinguish $C_A$ from $N_c$.

\subsection{Dimension 3: masses}
\label{sec:RGEsMasses}

Since neutrinos do not interact in QED or QCD, the Majorana neutrino mass does not receive an anomalous dimension in the LEFT at dimension five to any loop order, hence \\[-0.2cm]
\begin{equation}
	[ \dot M_\nu ]_2 = 0 \, .
\end{equation}
The RGEs for the other fermion mass matrices are
\begin{align}
	[ \dot M_e ]_2 &= - e^4 \q_e^2 \left( \q_e^2 (4n_e + 3) + 4 N_c (n_u \q_u^2 + n_d \q_d^2) \right) M_e \nn
		&\quad + 2 e^3 \q_e \left( \q_e^2 (4n_e + 7) + 4 N_c (n_u \q_u^2 + n_d \q_d^2) \right) ( L_{e\gamma}^\dagger M_e^\dagger M_e + M_e M_e^\dagger L_{e\gamma}^\dagger ) \nn
		&\quad + \frac{4}{3} e^3 \q_e \left( \q_e^2 (8n_e + 15) + 8 N_c (n_u \q_u^2 + n_d \q_d^2) \right) M_e  L_{e\gamma} M_e \nn
		&\quad - 40 e^3 \q_e^2 \Big( \q_e ( \<L_{e\gamma} M_e\> + \<L_{e\gamma}^\dagger M_e^\dagger\> ) \nn
			&\quad\qquad\qquad + N_c \q_u ( \<L_{u\gamma} M_u\> + \<L_{u\gamma}^\dagger M_u^\dagger\> ) + N_c \q_d ( \<L_{d\gamma} M_d\> + \<L_{d\gamma}^\dagger M_d^\dagger\> ) \Big) M_e \nn
		&\quad - 192 e^3 \q_e \Big( \q_e^2 \<M_e M_e^\dagger\> + N_c \q_u^2 \<M_u M_u^\dagger\> + N_c \q_d^2 \<M_d M_d^\dagger\> \Big) L_{e\gamma}^\dagger \, , \\
	[ \dot M_u ]_2 &= - e^4 \q_u^2 \left( 4n_e \q_e^2 + (4 N_c n_u + 3) \q_u^2 + 4 N_c n_d \q_d^2 \right) M_u \nn
		&\quad - g^2 C_F \left( 6 e^2 \q_u^2 + g^2 \left( 3(C_F+N_c) + 2(n_u+n_d) \right) \right) M_u \nn
		&\quad + 2 e \q_u \left( 7 ( g^2 C_F + e^2 \q_u^2) + 4 e^2 ( n_e \q_e^2 + N_c (n_u \q_u^2 + n_d \q_d^2) ) \right) ( L_{u\gamma}^\dagger M_u^\dagger M_u + M_u M_u^\dagger L_{u\gamma}^\dagger ) \nn
		&\quad + \frac{4}{3} e \q_u \left( 15 ( g^2 C_F + e^2 \q_u^2) +  8 e^2 (n_e \q_e^2 + N_c (n_u \q_u^2 + n_d \q_d^2) ) \right) M_u  L_{u\gamma} M_u \nn
		&\quad + 2 g C_F \left( 7 ( g^2 C_F + e^2 \q_u^2) + g^2 (19 N_c+2(n_u+n_d)) \right) ( L_{uG}^\dagger M_u^\dagger M_u + M_u M_u^\dagger L_{uG}^\dagger ) \nn
		&\quad + \left(  \frac{4}{3} g^3 C_F ( 15 C_F - 7 N_c + 4(n_u+n_d) ) + 20 g C_F e^2 \q_u^2 \right) M_u  L_{uG} M_u \nn
		&\quad - 40 e^3 \q_u^2 \Big( \q_e ( \<L_{e\gamma} M_e\> + \<L_{e\gamma}^\dagger M_e^\dagger\> ) \nn
			&\quad\qquad\qquad + N_c \q_u ( \<L_{u\gamma} M_u\> + \<L_{u\gamma}^\dagger M_u^\dagger\> ) + N_c \q_d ( \<L_{d\gamma} M_d\> + \<L_{d\gamma}^\dagger M_d^\dagger\> ) \Big) M_u \nn
		&\quad - 20 g^3 C_F \Big( \<L_{uG} M_u\> + \<L_{uG}^\dagger M_u^\dagger\> + \<L_{dG} M_d\> + \<L_{dG}^\dagger M_d^\dagger\> \Big) M_u \nn
		&\quad - 192 e^3 \q_u \Big( \q_e^2 \<M_e M_e^\dagger\> + N_c \q_u^2 \<M_u M_u^\dagger\> + N_c \q_d^2 \<M_d M_d^\dagger\> \Big) L_{u\gamma}^\dagger \nn
		&\quad - 96 g^3 C_F \Big( \<M_u M_u^\dagger\> + \<M_d M_d^\dagger\> \Big) L_{uG}^\dagger \, , \\
	[ \dot M_d ]_2 &= - e^4 \q_d^2 \left( 4n_e \q_e^2 + 4 N_c n_u \q_u^2 + (4 N_c n_d + 3) \q_d^2 \right) M_d \nn
		&\quad - g^2 C_F \left( 6 e^2 \q_d^2 + g^2 \left( 3(C_F+N_c) + 2(n_u+n_d) \right) \right) M_d \nn
		&\quad + 2 e \q_d \left( 7 ( g^2 C_F + e^2 \q_d^2) + 4 e^2 ( n_e \q_e^2 + N_c (n_u \q_u^2 + n_d \q_d^2) ) \right) ( L_{d\gamma}^\dagger M_d^\dagger M_d + M_d M_d^\dagger L_{d\gamma}^\dagger ) \nn
		&\quad + \frac{4}{3} e \q_d \left( 15 ( g^2 C_F + e^2 \q_d^2) +  8 e^2 (n_e \q_e^2 + N_c (n_u \q_u^2 + n_d \q_d^2) ) \right) M_d  L_{d\gamma} M_d \nn
		&\quad + 2 g C_F \left( 7 ( g^2 C_F + e^2 \q_d^2) + g^2 (19 N_c+2(n_u+n_d)) \right) ( L_{dG}^\dagger M_d^\dagger M_d + M_d M_d^\dagger L_{dG}^\dagger ) \nn
		&\quad + \left(  \frac{4}{3} g^3 C_F ( 15 C_F - 7 N_c + 4(n_u+n_d) ) + 20 g C_F e^2 \q_d^2 \right) M_d  L_{dG} M_d \nn
		&\quad - 40 e^3 \q_d^2 \Big( \q_e ( \<L_{e\gamma} M_e\> + \<L_{e\gamma}^\dagger M_e^\dagger\> ) \nn
			&\quad\qquad\qquad + N_c \q_u ( \<L_{u\gamma} M_u\> + \<L_{u\gamma}^\dagger M_u^\dagger\> ) + N_c \q_d ( \<L_{d\gamma} M_d\> + \<L_{d\gamma}^\dagger M_d^\dagger\> ) \Big) M_d \nn
		&\quad - 20 g^3 C_F \Big( \<L_{uG} M_u\> + \<L_{uG}^\dagger M_u^\dagger\> + \<L_{dG} M_d\> + \<L_{dG}^\dagger M_d^\dagger\> \Big) M_d \nn
		&\quad - 192 e^3 \q_d \Big( \q_e^2 \<M_e M_e^\dagger\> + N_c \q_u^2 \<M_u M_u^\dagger\> + N_c \q_d^2 \<M_d M_d^\dagger\> \Big) L_{d\gamma}^\dagger \nn
		&\quad - 96 g^3 C_F \Big( \<M_u M_u^\dagger\> + \<M_d M_d^\dagger\> \Big) L_{dG}^\dagger \, .
\end{align}

\subsection{Dimension 4: gauge couplings}

The two-loop running of the gauge couplings in the LEFT to dimension five is given by
\begin{align}
	\label{eq:eRGE}
	[ \dot e ]_2 &= - b^e_{1,0} e^5 - b^e_{1,1} e^3 g^2 \nn
		&\quad - 8 e^2 \bigg( 10 e^2 \q_e^3 \Big( \< L_{e\gamma} M_e \> + \< L_{e\gamma}^\dagger M_e^\dagger \> \Big) \nn
		&\qquad\qquad + N_c \q_u (9 g^2 C_F + 10 e^2 \q_u^2) \Big( \< L_{u\gamma} M_u \> + \< L_{u\gamma}^\dagger M_u^\dagger \> \Big) \nn
		&\qquad\qquad + N_c \q_d (9 g^2 C_F + 10 e^2 \q_d^2) \Big( \< L_{d\gamma} M_d \> + \< L_{d\gamma}^\dagger M_d^\dagger \> \Big) \nn
		&\qquad\qquad + e g N_c C_F \q_u^2 \Big( \< L_{uG} M_u \> + \< L_{uG}^\dagger M_u^\dagger \> \Big) \nn
		&\qquad\qquad + e g N_c C_F \q_d^2 \Big( \< L_{dG} M_d \> + \< L_{dG}^\dagger M_d^\dagger \> \Big) \bigg) \, , \\
	[ \dot g ]_2 &= - b^g_{1,0} g^5 - b^g_{1,1} g^3 e^2 \nn
		&\quad - g^2 \bigg( \left( 2 g^2 (20 C_F + 11 N_c) + 36 e^2 \q_u^2 \right) \Big( \< L_{uG} M_u \> + \< L_{uG}^\dagger M_u^\dagger \> \Big) \nn
		&\qquad\qquad + \left( 2 g^2 (20 C_F + 11 N_c) + 36 e^2 \q_d^2 \right) \Big( \< L_{dG} M_d \> + \< L_{dG}^\dagger M_d^\dagger \> \Big) \nn
		&\qquad\qquad + 4 e g \q_u \Big( \< L_{u\gamma} M_u \> + \< L_{u\gamma}^\dagger M_u^\dagger \> \Big) \nn
		&\qquad\qquad + 4 e g \q_d \Big( \< L_{d\gamma} M_d \> + \< L_{d\gamma}^\dagger M_d^\dagger \> \Big) \bigg) \, .
\end{align}
The two-loop running of the $\theta$-parameters is
\begin{align}
	\label{eq:thetaQEDRGE}
	[ \dot \theta_\mathrm{QED} ]_2 &= \frac{128\pi^2 i}{e} \bigg( 10 e^2 \q_e^3 \Big( \< L_{e\gamma} M_e \> - \< L_{e\gamma}^\dagger M_e^\dagger \> \Big) \nn
		&\qquad\qquad + N_c \q_u (9 g^2 C_F + 10 e^2 \q_u^2) \Big( \< L_{u\gamma} M_u \> - \< L_{u\gamma}^\dagger M_u^\dagger \> \Big) \nn
		&\qquad\qquad + N_c \q_d (9 g^2 C_F + 10 e^2 \q_d^2) \Big( \< L_{d\gamma} M_d \> - \< L_{d\gamma}^\dagger M_d^\dagger \> \Big) \nn
		&\qquad\qquad + e g N_c C_F \q_u^2 \Big( \< L_{uG} M_u \> - \< L_{uG}^\dagger M_u^\dagger \> \Big) \nn
		&\qquad\qquad + e g N_c C_F \q_d^2 \Big( \< L_{dG} M_d \> - \< L_{dG}^\dagger M_d^\dagger \> \Big) \bigg) \, , \\
	\label{eq:thetaQCDRGE}
	[ \dot \theta_\mathrm{QCD} ]_2 &= \frac{16\pi^2 i}{g} \bigg( \left( 2 g^2 (20 C_F + 3 N_c) + 36 e^2 \q_u^2 \right) \Big( \< L_{uG} M_u \> - \< L_{uG}^\dagger M_u^\dagger \> \Big) \nn
		&\qquad\qquad + \left( 2 g^2 (20 C_F + 3 N_c) + 36 e^2 \q_d^2 \right) \Big( \< L_{dG} M_d \> - \< L_{dG}^\dagger M_d^\dagger \> \Big) \nn
		&\qquad\qquad + 4 e g \q_u \Big( \< L_{u\gamma} M_u \> - \< L_{u\gamma}^\dagger M_u^\dagger \> \Big) \nn
		&\qquad\qquad + 4 e g \q_d \Big( \< L_{d\gamma} M_d \> - \< L_{d\gamma}^\dagger M_d^\dagger \> \Big) \bigg) \, .
\end{align}

\subsection{Dimension 5: dipole operators}

At dimension five, the $\Delta L = 2$ neutrino dipole operator obtains with our normalization an anomalous dimension only through the photon field,
\begin{equation}
	[ \dot L_{\nu\gamma} ]_2 = - (b^e_{1,0} e^4 + b^e_{1,1} e^2 g^2) L_{\nu\gamma} \, .
\end{equation}
The RGEs for the remaining dipole operators are
\begin{align}
	[ \dot L_{e\gamma} ]_2 &=  - \left[  (b^e_{1,0} e^4 + b^e_{1,1} e^2 g^2) + \frac{e^4 \q_e^2}{9} \left( 459 \q_e^2 + 116\left( n_e \q_e^2 + N_c (n_u \q_u^2 + n_d \q_d^2) \right) \right) \right] L_{e\gamma} \, , \\
	[ \dot L_{u\gamma} ]_2 &=  - \bigg[  (b^e_{1,0} e^4 + b^e_{1,1} e^2 g^2) + \frac{e^4 \q_u^2}{9} \left( 459 \q_u^2 + 116\left( n_e \q_e^2 + N_c (n_u \q_u^2 + n_d \q_d^2) \right) \right) \nn
		&\quad\qquad  + \frac{g^2 C_F}{9} \left( 630 e^2 \q_u^2 + 26 g^2 ( n_u + n_d ) + 171 g^2 C_F - 257 g^2 N_c \right) \bigg] L_{u\gamma} \nn
		&\quad  - \frac{4 e g C_F \q_u}{9} \left( 72 e^2 \q_u^2 + 8 g^2 ( n_u + n_d ) + 72 g^2 C_F - 95 g^2 N_c \right) L_{uG} \, , \\
	[ \dot L_{d\gamma} ]_2 &=  - \bigg[  (b^e_{1,0} e^4 + b^e_{1,1} e^2 g^2) + \frac{e^4 \q_d^2}{9} \left( 459 \q_d^2 + 116\left( n_e \q_e^2 + N_c (n_u \q_u^2 + n_d \q_d^2) \right) \right) \nn
		&\quad\qquad  + \frac{g^2 C_F}{9} \left( 630 e^2 \q_d^2 + 26 g^2 ( n_u + n_d ) + 171 g^2 C_F - 257 g^2 N_c \right) \bigg] L_{d\gamma} \nn
		&\quad  - \frac{4 e g C_F \q_d}{9} \left( 72 e^2 \q_d^2 + 8 g^2 ( n_u + n_d ) + 72 g^2 C_F - 95 g^2 N_c \right) L_{dG} \, , \\
	[ \dot L_{uG} ]_2 &=  - \bigg[  (b^g_{1,0} g^4 + b^g_{1,1} g^2 e^2) + \frac{e^4 \q_u^2}{9} \left( 171 \q_u^2 + 52\left( n_e \q_e^2 + N_c (n_u \q_u^2 + n_d \q_d^2) \right) \right) \nn
		&\quad\qquad  - 2 e^2 g^2 \q_u^2 (12 N_c - 35 C_F) \nn
		&\quad\qquad  - \frac{g^4}{36} \left(643 N_c^2 - 932 - \frac{459}{N_c^2} - 4 (58 C_F - 13 N_c) (n_u + n_d) \right) \bigg] L_{uG} \nn
		&\quad  - \left[ 8 e g^3 \q_u (4 C_F - N_c) + \frac{32 e^3 g \q_u}{9} \left(9 \q_u^2 + 2 \left(n_e \q_e^2 + N_c ( n_u \q_u^2 + n_d \q_d^2 ) \right) \right) \right] L_{u\gamma} \, , \\
	[ \dot L_{dG} ]_2 &=  - \bigg[  (b^g_{1,0} g^4 + b^g_{1,1} g^2 e^2) + \frac{e^4 \q_d^2}{9} \left( 171 \q_d^2 + 52\left( n_e \q_e^2 + N_c (n_u \q_u^2 + n_d \q_d^2) \right) \right) \nn
		&\quad\qquad  - 2 e^2 g^2 \q_d^2 (12 N_c - 35 C_F) \nn
		&\quad\qquad  - \frac{g^4}{36} \left(643 N_c^2 - 932 - \frac{459}{N_c^2} - 4 (58 C_F - 13 N_c) (n_u + n_d) \right) \bigg] L_{dG} \nn
		&\quad  - \left[ 8 e g^3 \q_d (4 C_F - N_c) + \frac{32 e^3 g \q_d}{9} \left(9 \q_d^2 + 2 \left(n_e \q_e^2 + N_c (n_u \q_u^2 + n_d \q_d^2) \right) \right) \right] L_{d\gamma} \, .
\end{align}

	\clearpage

	\addcontentsline{toc}{section}{\numberline{}References}
	\bibliographystyle{utphysmod}
	\bibliography{Literature}
	
\end{document}